\documentclass[aps,prl,twocolumn,superscriptaddress,showpacs]{revtex4-1}
\usepackage[version=3]{mhchem}
\usepackage{graphicx}
\usepackage{siunitx}

\usepackage{hyperref}

\usepackage{url}

\usepackage{color,soul}

\usepackage{booktabs}
\usepackage[draft]{changes}
\usepackage{ctable}

\usepackage{miller}

\usepackage{cleveref}

\newcommand{\dto}{Dy$_2$Ti$_2$O$_7$}

\newcommand{\deltaL}{\Delta L/L}

\begin{document}

%\preprint{}

\title{Extremely slow non-equilibrium monopole dynamics in classical spin ice}


\author{T. St\"oter}
%\thanks{}
\affiliation{Institute for Solid State and Materials Physics and W\"urzburg-Dresden Cluster of Excellence $ct.qmat$, TU Dresden, 01062 Dresden,Germany}
\affiliation{Hochfeld-Magnetlabor Dresden (HLD-EMFL), Helmholtz-Zentrum
Dresden-Rossendorf, 01328 Dresden, Germany}

\author{M. Doerr}
\affiliation{Institute for Solid State and Materials Physics and W\"urzburg-Dresden Cluster of Excellence $ct.qmat$, TU Dresden, 01062 Dresden, Germany}
\author{S. Granovsky}
\affiliation{Institute for Solid State and Materials Physics and W\"urzburg-Dresden Cluster of Excellence $ct.qmat$, TU Dresden, 01062 Dresden, Germany}
\affiliation{Faculty of Physics, M.~V. Lomonosov Moscow State University, Moscow, 119991, Russia}
\author{M. Rotter}
\affiliation{McPhase Project, 01159 Dresden, Germany}
\author{S.~T.~B. Goennenwein }
\affiliation{Institute for Solid State and Materials Physics and W\"urzburg-Dresden Cluster of Excellence $ct.qmat$, TU Dresden, 01062 Dresden, Germany}
\author{S. Zherlitsyn}
\affiliation{Hochfeld-Magnetlabor Dresden (HLD-EMFL), Helmholtz-Zentrum
Dresden-Rossendorf, 01328 Dresden, Germany}
\author{O. A. Petrenko}
\author{G. Balakrishnan}
\affiliation{University of Warwick, Department of Physics, Coventry CV4 7AL,
United Kingdom}
\author{H. D. Zhou}
\affiliation{University of Tennessee, Department of Physics and Astronomy,
Knoxville, Tennessee 37996-1200, USA}
\affiliation{National High Magnetic Field Laboratory, Florida State University,
Tallahassee 32306-4005, USA}
\author{J. Wosnitza}
\affiliation{Institute for Solid State and Materials Physics and W\"urzburg-Dresden Cluster of Excellence $ct.qmat$, TU Dresden, 01062 Dresden,Germany}
\affiliation{Hochfeld-Magnetlabor Dresden (HLD-EMFL), Helmholtz-Zentrum
Dresden-Rossendorf, 01328 Dresden, Germany}


\date{\today}

\begin{abstract}
We report on the non-equilibrium monopole dynamics in the classical spin ice \dto{} detected
by means of high-resolution magnetostriction measurements.
Significant lattice changes occur at the transition from the kagome-ice to the
	saturated-ice phase, visible in the longitudinal and transverse
	magnetostriction.
A hysteresis opening at temperatures below \SI{0.6}{\kelvin} suggests a first-order
    transition between the kagome and saturated state.
Extremely slow lattice relaxations, triggered by changes of the magnetic field, were
	observed. These lattice-relaxation effects result
	from non-equilibrium monopole formation or annihilation processes.
The relaxation times extracted from our experiment are in good agreement with theoretical predictions with
	decay constants of the order of $10{^4}$~s at \SI{0.3}{\kelvin}.
\end{abstract}

%\pacs{75.40.Cx, 75.40.Gb, 81.10.-h}
%\keywords{}

\maketitle

%\section{Introduction}

Magnetically frustrated materials are the subject of intense research due to inherently
competing interactions, large ground-state degeneracy, the appearance of exotic states, such as
spin-ice and spin-liquid phases~\citep{moessner_pt_2006,balents_spin_2010}, deconfined fractionalized excitations
(magnetic monopoles)~\citep{castelnovo_magnetic_2008},
non-stationary processes~\citep{slobinsky_unconventional_2010,erfanifam_intrinsic_2011, erfanifam_ultrasonic_2014},
and unusual spin dynamics~\citep{borzi_nc_2016,kaiser_2015,paulsen_nc_2019,paulsen_np_2016,paulsen_far-from-equilibrium_2014,yaraskavitch_spin_2012,
giblin_prl_2018}. However, an analysis of the long-term non-equilibrium processes at low fields is given only in one paper ~\citep{paulsen_far-from-equilibrium_2014}.

Prominent examples of frustrated magnetic systems are the pyrochlore oxides \ce{$A$2$B$2O7},
    with a trivalent rare-earth ion \ce{$A$^3+} and a tetravalent ion \ce{$B$^4+}.
    Especially in the pyrochlores with \ce{Dy} or \ce{Ho} on the $A$ site,
	many different exotic states have been investigated.
The single-ion ground state can be treated as an effective spin-half state
	\citep{rau_magnitude_2015}.
The two-ion interaction is very well described by the dipolar spin-ice model
	\citep{den_hertog_dipolar_2000}
	which includes dipolar and exchange interaction
	that result in an effective ferromagnetic
	nearest-neighbor interaction.
These interactions, together with strong magnetic anisotropy due to crystal-electric-field (CEF) effects, favor the highly degenerate spin-ice configuration:
	two spins point into and two spins point out of each ``2-in-2-out'' tetrahedron~\citep{bramwell_spin_2001}.
The excitations of this arrangement to ``3-in-1-out'' or ``1-in-3-out''can be interpreted as the creation of
	magnetic monopole-antimonopole pairs~\cite{castelnovo_magnetic_2008}.
%\hl{This ground state is characterized by a residual entropy of $1/2 \ln{3/2}R$
%	{\citep{ramirez_zero-point_1999}}.}

The study of thermally activated spin dynamics in \dto{} via ac-susceptibility
    or magnetization is the focus of numerous publications
    ~{\citep{matsuhira_novel_2001, snyder_quantum-classical_2003,
	snyder_low-temperature_2004, matsuhira_spin_2011,
	yaraskavitch_spin_2012, matsuhira_low_2000,
	snyder_how_2001, ehlers_dynamical_2003, ehlers_evidence_2004, petrenko_mag_2011,
	quilliam_dynamics_2011}} because these slow dynamic effects are directly connected to the
    kinetics and interactions of these monopoles. In detail, a sharp increase of the relaxation
    times at temperatures below 1~K was reported and attributed
    to the Coulomb-gas character of charged particles (monopoles) forming a network of `Dirac strings'. This
    blocks the monopoles in metastable states~{\citep{jaubert_magnetic_2009, jaubert_magnetic_2011, giblin_relax_2011}}.
    The temperature-dependent non-equilibrium dynamics was analyzed theoretically in ref.~\cite{castelnovo_thermal_2010}.
Field-dependent magnetic investigations which also resolve the monopole dynamics in the ``3-in-1-out'' or ``1-in-3-out''
    kagome state or in the saturated spin-ice in relation to the crystal lattice are, although of great interest~\cite{mostame_tunable_2014}, rather rare and, therefore, are the main topic of this report.

%In several investigations of {\dto{}} and {\hto{}} via ac-susceptibility
%	measurements{\citep{matsuhira_novel_2001, snyder_quantum-classical_2003,
%	snyder_low-temperature_2004, matsuhira_spin_2011,
%	yaraskavitch_spin_2012, matsuhira_low_2000,
%	snyder_how_2001, ehlers_dynamical_2003, ehlers_evidence_2004,
%	quilliam_dynamics_2011}} thermally-activated spin dynamics have been
%	found that are attributed to the thermal excitation of
%	monopoles{\citep{jaubert_magnetic_2011}}.
%The magnetization and specific heat measurements of the spin-ices
%	show characteristic behavior like a liquid-gas transition at sub-kelvin
%	temperatures in field-dependent measurements%
%	{\citep{sakakibara_observation_2003, krey_first_2012,
%	petrenko_magnetization_2003}}
%	or a Schottky-like anomaly in temperature-dependent measurements at
%	zero field{\citep{hiroi_specific_2003,
%	higashinaka_low_2004,bramwell_spin_correlation_2001}}, respectively.

Several theoretical~\citep{yamashita_spin-driven_2000,
	tchernyshyov_order_2002,
	tchernyshyov_spin-peierls_2002,richter_magnetic-field_2004,
	jia_lattice-coupled_2005,penc_symmetry_2007,terao_distortion_2007,
	saunders_structural_2008,curnoe_structural_2008,
	aoyama_spin-lattice-coupled_2016} and experimental
	studies~\citep{mirebeau_spin_2004,erfanifam_intrinsic_2011,
	erfanifam_ultrasonic_2014} have investigated the
	magnetoelastic coupling in pyrochlore systems.
As frustration is highly dependent on the symmetry of the lattice, distorting
	the lattice can relieve the frustration.
The spin-ice state was shown to be stable under hydrostatic pressure
	\citep{mirebeau_spin_2004}.
Likewise lifting the degeneracy of the frustrated state by a magnetic field
	could also influence the lattice.

%Applying a field in \hkl[1 1 1] direction triggers two transitions: First, a
%	crossover from the spin ice phase to the so called kagome ice phase;
%	second, a transition from the kagome ice state to the saturated ice
%	state, that is first order below a critical point at
%	\SI{0.6}{\K}\cite{sakakibara_observation_2003}.
In ultrasound measurements on \dto{}, dramatic
	non-equilibrium effects were found: thermal runaway associated with
	monopole avalanches \cite{erfanifam_intrinsic_2011}.
In the magnetization, these kinds of avalanches also exist~\cite{jackson_dynamic_2014}.
Apart from these short-time-scale effects, an increase of the time scale of the
	internal dynamics has been observed in various measurements
	on spin ice, such as in ac susceptibility \citep{matsuhira_novel_2001,
	snyder_quantum-classical_2003,snyder_low-temperature_2004,
	matsuhira_spin_2011,yaraskavitch_spin_2012, takatsu_ac_2013,
	matsuhira_low_2000,snyder_how_2001,ehlers_dynamical_2003,
	ehlers_evidence_2004,quilliam_dynamics_2011}, magnetization
	\citep{matsuhira_spin_2011,paulsen_far-from-equilibrium_2014,
	jackson_dynamic_2014}, the magnetocaloric effect
	\citep{orendac_magnetocaloric_2007}, thermal conductivity
	\citep{li_low-temperature_2015}, and heat capacity
	\citep{pomaranski_absence_2013}. Most of these measurements were performed quenching
    the sample from low fields to zero.
From theory, another kind of non-equilibrium effect when quenching
	the field from the monopole-rich saturated-ice or kagome-ice phase
	towards the spin-ice phase with only a few monopoles was suggested.
Both, short- and long-time annihilation processes have been proposed in the theory work of
    Mostame~\textit{et al.}~\cite{mostame_tunable_2014}. In experiment, a slow relaxation
    process has been reported to occur~\cite{paulsen_far-from-equilibrium_2014}.
    These investigations, however, were performed only at very low magnetic fields.
    In our studies, on the other hand, we investigate in detail non-equilibrium processes
    at higher fields, in the kagome-ice state, and, thereby, validate the theoretically proposed
    behavior~\cite{mostame_tunable_2014}.

In detail, we observed extremely slow lattice-relaxation processes which can be directly connected to
    the generation and annihilation of monopole/antimonopole pairs in magnetic field.
These macroscopic lattice changes allow us a direct insight into the microscopic spin dynamics.
    Via magnetoelastic coupling we can indirectly resolve the relevant minute magnetization changes
    with extraordinary high resolution.
\\
%\section{Experimental}
%\label{sec:experimental}

Single crystals of \dto{} were grown by the floating-zone technique~\citep{balakrishnan_single_1998}
    and oriented along the required crystallographic axes using x-ray Laue diffraction.
The crystals used for the experiments were oblate cuboids of dimensions $\sim$3~$\times$~2~$\times$~1 mm$^3$, for which a
    demagnetization factor of about 0.7 for fields along the shortest dimension was deduced.

For the measurements, we used a capacitive dilatometer~\cite{kuchler_compact_2012} that was mounted on
    a probe placed in a sorb-pumped \ce{^3He}-cryostat reaching temperatures down to 0.3~K.
    For accurate monitoring and control of the dilatometer and sample temperature we used
    a Cernox and a \ce{RuO2} thermometer (below
	\SI{1}{\kelvin}). Both thermometers were attached to the dilatometer cell close to the sample.

\begin{figure}
	\centering
	\includegraphics[width=\linewidth]{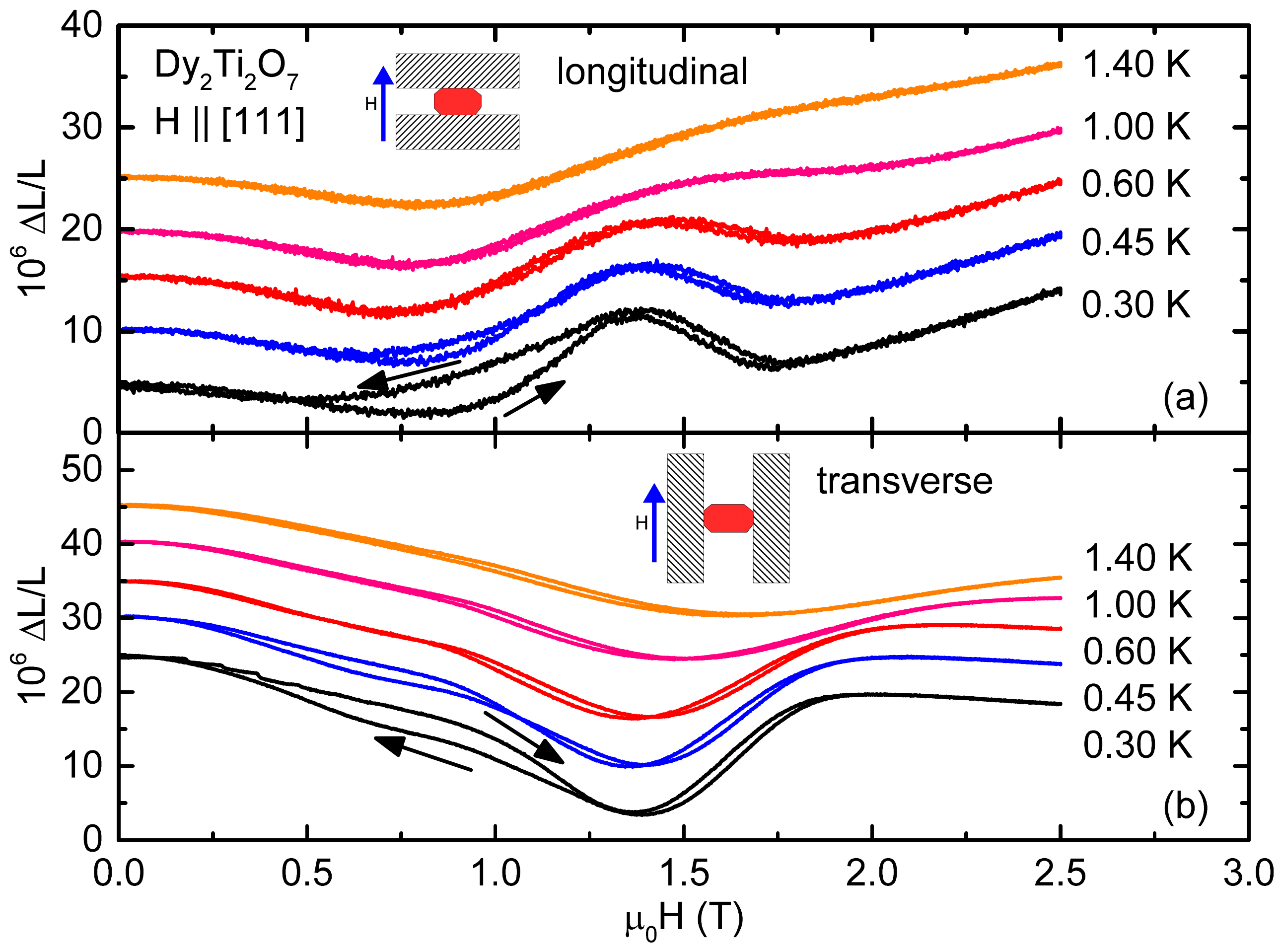}
	\caption{Field dependence of the relative sample-length change parallel and perpendicular to a
		magnetic field applied along the [111] direction at various temperatures.  The black arrows
        indicate the field-sweep direction. The sketches show
        how the sample (red), the magnetic field (blue arrow), and the capacitor plates (grey)
        are arranged. The curves are offset for clarity.}
	\label{fig:magnetostriction_trans_long}
\end{figure}

First, we focus on the (static) low-temperature magnetostriction curves displayed
    in Fig.~\ref{fig:magnetostriction_trans_long}.
Sweeping the field up at lowest temperature of \SI{0.3}{\kelvin} in longitudinal geometry,
    the crystal contracts at first until it reaches a minimum at around \SI{0.8}{\tesla}, then
	expands up to about \SI{1.3}{\tesla}. Upon further increasing the field,
    it contracts again reaching a local minimum and turning again to expand linearly with increasing
    field with a slope of about \SI{1d-5}{\tesla^{-1}}. Sweeping further up in field, there is no sign of saturation
    up to \SI{10}{\tesla} (not shown). This increase of the longitudinal magnetostriction is due to the mixing in of
    higher CEF terms, as simulations with the program package \textsc{McPhase}~\cite{rotter_mcphase_nodate} have shown.
%, as is suggested by previous measurements~\cite{kolland_anisotropic_2013}.

%The hysteresis, observed in the kagome-ice region, hints the transitions in the spin system to be of first order.
    The magnetostriction in transversal geometry is of roughly the oppositive character. All
    magnetostrictive effects are on the order of $10^{-5}$, well above the noise level of about $10^{-7}$.
%The normal noise level of $\approx 10^{-7}$ is probably caused by thermal instabilities.
    There is a distinct anomaly at about \SI{1.3}{\tesla} corresponding to the first-order transition between the kagome and ``3-in-1-out'' state~\cite{hiroi_specific_2003} in good agreement with our own simulations.
    In the following, we will only discuss the results obtained in longitudinal geometry. The transverse magnetostriction shows similar relaxation phenomena as discussed below.

In the kagome-ice region, we observe a clear hysteresis between about 0.5 and 1.2~T below 0.6~K (Fig.~\ref{fig:magnetostriction_trans_long}).
    It must be assumed that extensive dynamic effects
    are present in this area and that
    the sample is not in an equilibrium state in this range.

We investigated this latter point in more detail by probing the lattice relaxation and non-equilibrium
    spin dynamics using clearly specified protocols. Before each measurement sequence, we prepared the
    sample in a well-defined state. After oscillating the field in order to demagnetize the magnet,
    the sample was cooled down to the measurement temperature and thermally stabilized
    for \SI{30}{\minute} at \SI{0}{\tesla}. To illustrate, the following sequence gives an example how
    we performed our measurements. We increased, at constant temperature, the magnetic field at a fast
    sweep rate of \SI{1}{\tesla \per \minute} in steps between 0.1 and \SI{0.5}{\tesla} up to the final
    field and decreased stepwise afterwards (with the same sweep rate). After that, we monitored the
    lattice relaxation using our dilatometer for \SI{1}{\hour} or longer at constant field and temperature.

\begin{figure}
	\centering
	\includegraphics[width=\linewidth]{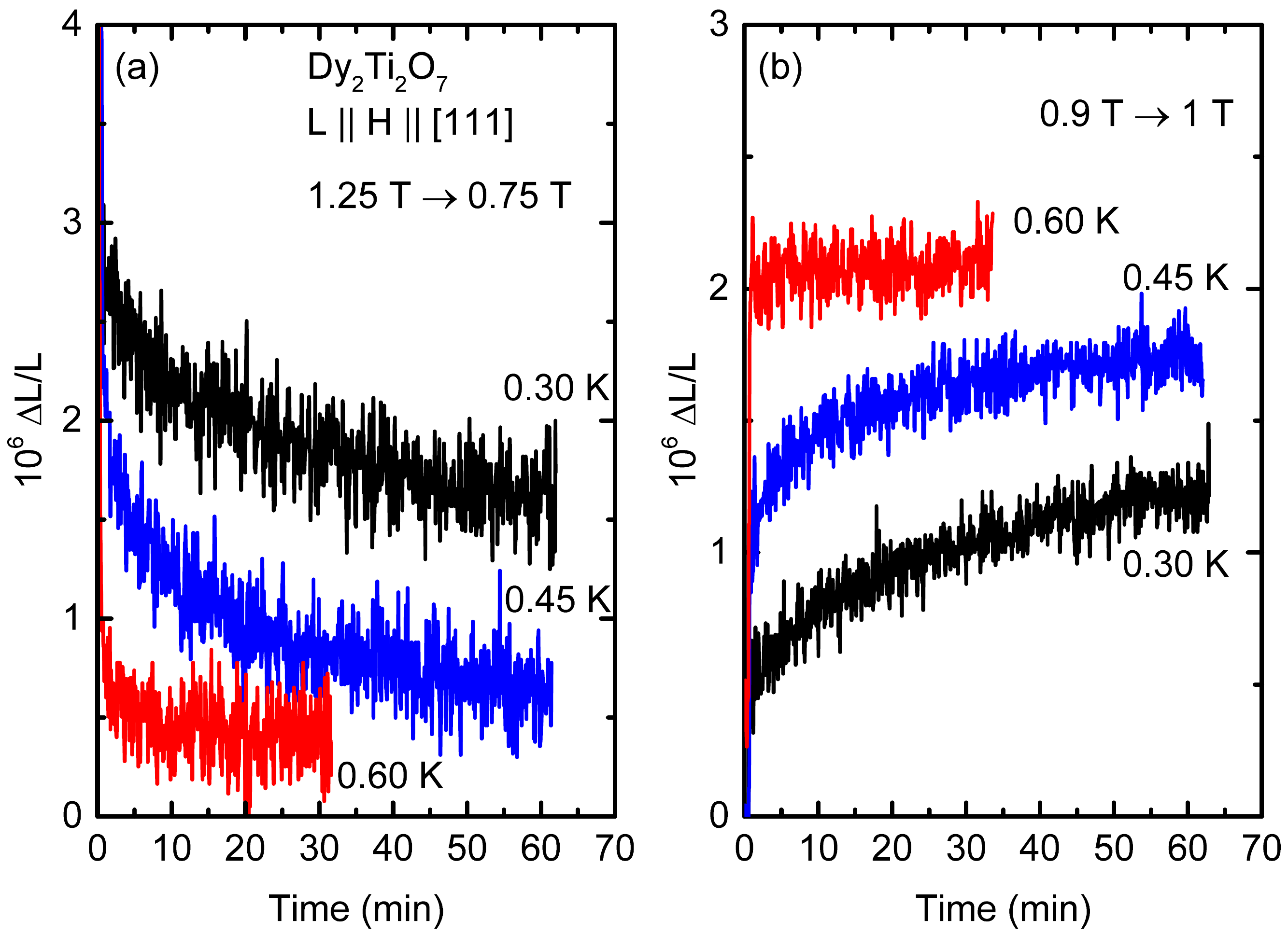}
	\caption{Typical time dependence of the relative sample length of \dto{}
		in longitudinal geometry $H \| \deltaL{}$
		%(\SI{0.97}{\tesla} to \SI{0.47}{\tesla}
		%including demagnetizing effects)
		at various temperatures
		(a) sweeping the field quickly from \SI{1.25} down to
		\SI{0.75}{\tesla}
		and (b) sweeping from \SI{0.90} up to
		\SI{1.00}{\tesla}.
		\label{fig:relaxationdto}
		}
\end{figure}
%For \dto{}, I used sequence A, stepping the field up and down in steps of
%	\SI{0.5}{\tesla} and waiting time of \SI{30}{} or \SI{60}{\minute}, to
%	obtain an overview of the occurrence of lattice-relaxation effects.
Some typical relaxation data for \dto{} in longitudinal geometry
	are shown in Fig.~\ref{fig:relaxationdto} for various temperatures.
These plots display the changes in sample length as a function of time;
	Figure~\ref{fig:relaxationdto}(a) shows the data after rapid (1~T/min) reduction of the magnetic field,
	Fig.~\ref{fig:relaxationdto}(b) those after field increase.
%The noise level in the transverse measurement is smaller than
%	in the longitudinal geometry because of the sample geometry with larger
%	width and small thickness.
While decreasing the field [Fig.~\ref{fig:relaxationdto}(a)], the lattice contracts quickly, but even after we stop
    the field sweep, the lattice still continues to contract very slowly. For \SI{0.3}and \SI{0.45}{\kelvin}, even after \SI{1}{\hour} no steady state is reached. At \SI{0.6}{\kelvin}, the lattice changes stop after
	about \SI{600}{\second}, $i.e.$, the lattice relaxation is slower at lower temperatures.
    We found corresponding behavior for increasing fields [Fig.~\ref{fig:relaxationdto}(b)]. During the sweeps, the lattice expands quickly and continues to do so very slowly even after stopping the sweep. Again, we observe much longer time scales of the lattice relaxation the lower the temperature.

\begin{figure}
	\centering
	\includegraphics[width=\linewidth]{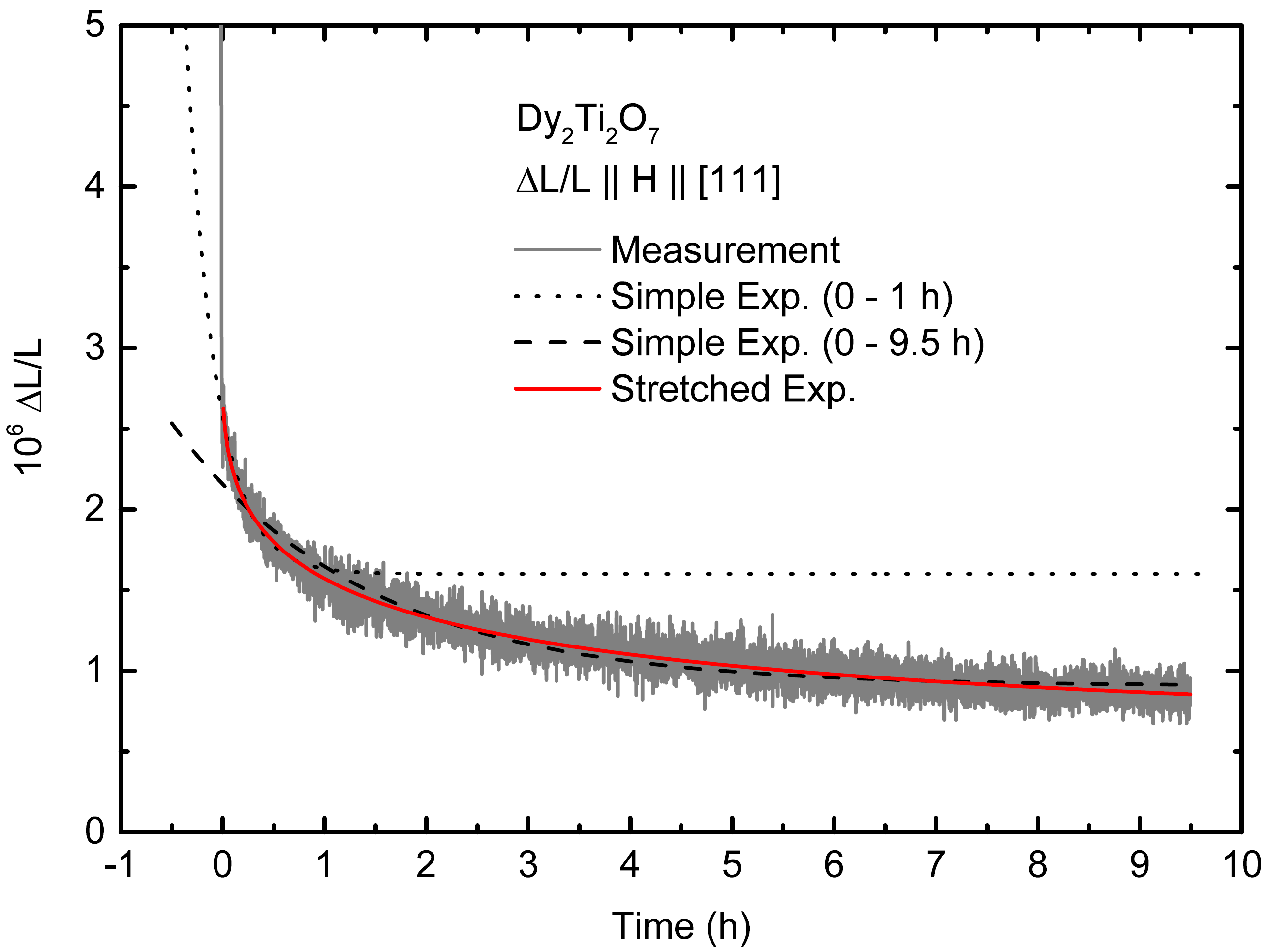}
	\caption{Long-time change of the sample length after
	a quick field sweep from \SI{1.25}{} to \SI{0.75}{\tesla} and fit lines using different models to describe the data.
	\label{fig:longrelaxation}}
\end{figure}

In order to analyze the data quantitatively and to obtain more insight into the non-equilibrium dynamics involved, we need some model
	to describe the relaxation time.
    It seems reasonable that some exponential decay law should be adequate to describe the data, but the exact form is not $a~priori$
	known.
    Therefore, we performed a long-time relaxation measurement (10~h, Fig.~\ref{fig:longrelaxation}) to better resolve what kind
    of decay law describes the data best. The field was swept from \SI{1.25} to \SI{0.75}{\tesla} at \SI{0.3}{\kelvin}.
%In subsequent studies we investigated the relaxation in several steps.
%under various conditions.

%\paragraph{Very-long-time relaxation}

%In order to get an idea what kind of decay law was adequate to fit the data
%	best,
%	we performed an experiment with a very long time of over \SI{9}{\hour}
%	after a quick field sweep to monitor the time dependence of the
%	relative length change of the sample.
%The field was swept from \SI{1.25}{\tesla} to \SI{0.75}{\tesla} at a temperature
%	of \SI{0.3}{\kelvin}; this is shown in \cref{fig:longrelaxation}.
Possible models would be a simple exponential decay with one relaxation time

\begin{align}
	(\deltaL{})(t) & = (\deltaL{})_{\infty} + A e^{-(t-t_0)/\tau}\,,
	\label{eq:simple_exp}
\end{align}
or a stretched-exponential decay
\begin{align}
	(\deltaL{})(t) & = (\deltaL{})_{\infty} +
		A e^{-\left((t-t_0)/\tau\right)^\beta}\,,
	\label{eq:stretched_exp}
\end{align}
where $\tau$ is the relaxation time, $\beta$ an exponent describing the relaxation-time
    distribution, $(\Delta L/L)_\infty$ the relative sample-length change for $t \rightarrow \infty$,
    and $(\Delta L/L)_\infty + A$ gives the value of $(\Delta L/L)$ at $t = t_0$, with $t_0$ usually set to zero.
    The data were described using the free parameters $A$, $(\Delta L/L)_\infty$, and $\tau$ for both equations and additionally $\beta$ for the stretched-exponential fit.
%The parameters were found by a least-squares algorithm.
%In \cref{fig:longrelaxation} the long-time relaxation data are shown together
%	with a stretched-exponential, a simple exponential describing the data
%	in the first hour after the field sweep and a simple exponential
%	describing all the data after the field sweep.
We found that simple exponential decays using Eq.~(1) deviate significantly from the
	data, either at the beginning or at the end of the relaxation process (dashed and dotted lines in Fig.~\ref{fig:longrelaxation},
    respectively). On the other hand, the stretched exponential fit provides a much better match
    to the data in the whole time range (red line in Fig.~\ref{fig:longrelaxation}). The stretched-exponential model describes the data best for $\beta = 0.4$ and $\tau = \SI{5500}{\second}$ .

A model with two relaxation times $\tau_S$ and $\tau_L$ to distinguish between free and bound monopoles,
    as proposed in~\cite{matsuhira_spin_2011}, describes our data less well, although even more parameters are used.
    Indeed, the stretched-exponential model is better motivated:
    Firstly, we observed that any exponential-decay fit saturates too early and should be
    stretched over longer times.
    Secondly, applying model (2) is justified by the existence of the mutual interactions of the magnetic monopoles
    as suggested by theoretical investigations of the spin-ice model
	\citep{mostame_tunable_2014,revell_evidence_2013}.
Indeed, this model leads to a distribution of relaxation times.
In particular, the formation of incontractable monopole-antimonopole pairs
	might lead to a slowing down of the dynamics.

%Due to the high time consumption only one very-long-time measurement was done
%	and the other measurements had to be performed on shorter timescales.
%We could still get good estimates of the relaxation time $\tau$ by fixing
%	$\beta$ to the well-determined value of $0.4$ also for other
%	experiments.

%\paragraph{Quenches to and within the kagome ice phase -- Sequence A and B}

%\begin{figure}
%	\centering
%	\includegraphics[width=\linewidth]{./DTORelaxationTimes_StretchedExp_long.pdf}
%	\caption{Field dependence of the relaxation times of \dto{} at various temperatures.
%	\label{fig:relaxation_dto_stretched}}
%\vspace{0.5cm}
%	\centering
%	\includegraphics[width=\linewidth]{./DTORelaxationTimes_Stretched_temperature.pdf}
%	\caption{Temperature dependence of the relaxation times of \dto{} at various fields.
%	\label{fig:relaxation_dto_temperature}}
%\end{figure}

\begin{figure}
	\centering
	\includegraphics[width=\linewidth]{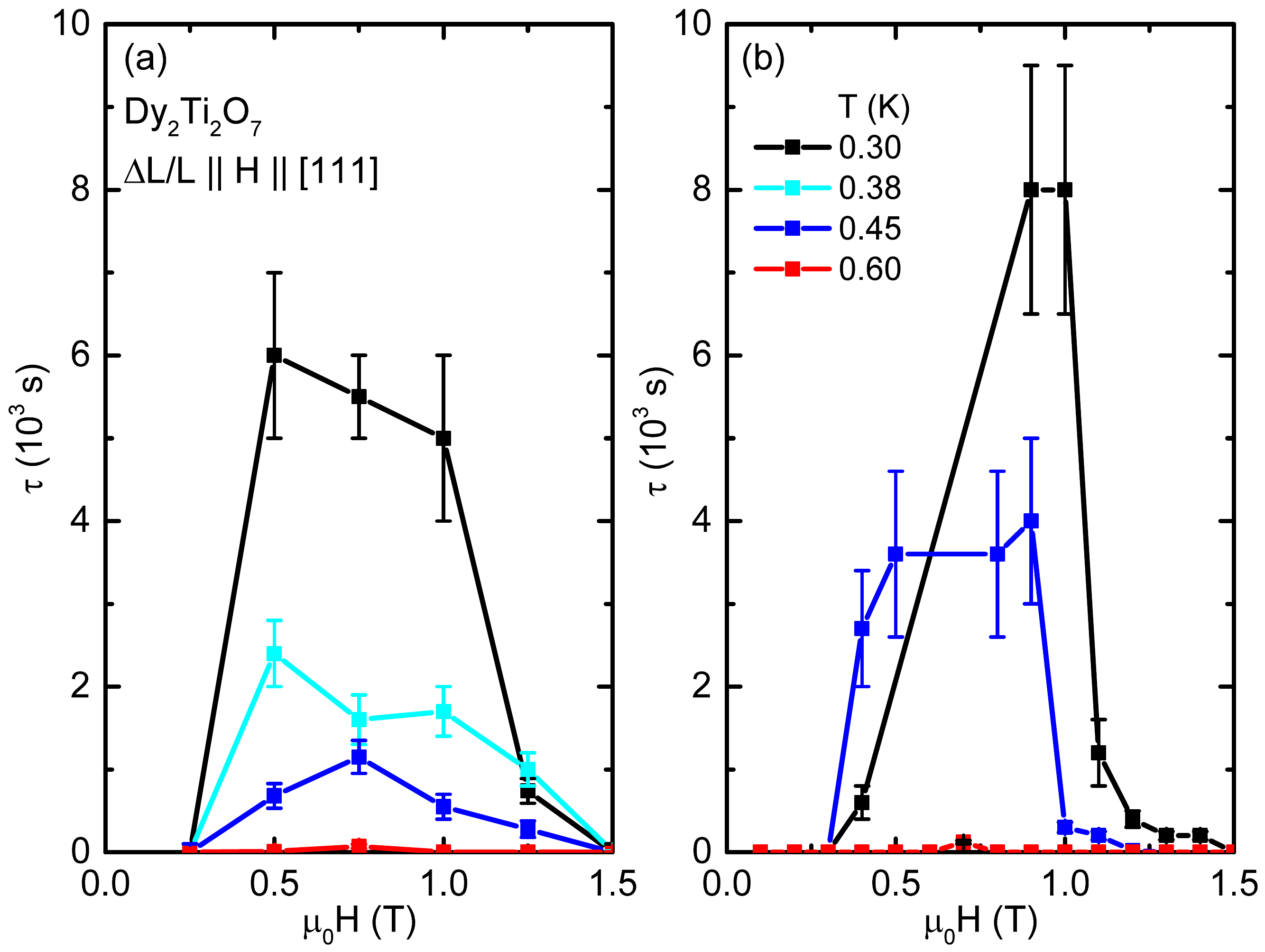}
	\caption{Field dependence of the relaxation times of \dto{} at various magnetic fields in
    longitudinal geometry $H \| \deltaL{}$\\
    after (a) reducing and (b) increasing the magnetic field quickly.
	\label{fig:relaxationtimedto}}
\end{figure}

%Sequence A in the down-stepping phase and sequence B set the system to the
%	saturated-ice phase in the beginning and followed by a sweeping the
%	field down in several steps from there.
%That means that the relaxation times at \SI{0.5}{\tesla} and \SI{0.75}{\tesla}
%	are not reached in one step from the saturated-ice phase.
%We obtained the relaxation times from these sequences by fitting the
%	stretched exponential function using \cref{eq:stretched_exp} with
%	$\beta = 0.4$ (obtained from the very-long-time relaxation experiment)
%	and $t_0$ fixed to the end of the field sweep.
In the following, we fixed $\beta$ to $0.4$ and used Eq.~(2)
   to describe our lattice-relaxation data (such as those shown in Fig.~\ref{fig:relaxationdto}).
The error bars were estimated by varying $\beta$ by $\pm 0.2$ and checking whether the
    data could be described with another set of $A$ and $\tau$.
%The difference to the largest $\tau_0$ for which this was possible is the error
%	of $\tau_0$.

\Cref{fig:relaxationtimedto} shows the field dependence of the extracted
	relaxation times of \dto{} after rapidly decreasing [Fig.~\ref{fig:relaxationtimedto}(a)] and increasing the field [Fig.~\ref{fig:relaxationtimedto}(b)].
The longest relaxation times were measured in the field region between
	\SI{0.5}{} and \SI{1.0}{\tesla}.
Non-equilibrium dynamics could not be observed at fields below \SI{0.2}{\tesla} or
	above \SI{1.2}{\tesla}.
The time scales of the relaxation are up to hours at \SI{0.3}{\kelvin} and a few minutes
	at \SI{0.6}{\kelvin}. Above \SI{0.7}{\kelvin}, no relaxation could be found.
This matches approximately the spin-freezing temperature of about
	\SI{0.6}{\kelvin} \citep{fukazawa_magnetic_2002}.
The extremely slow spin dynamics, therefore, is only present in the kagome-ice region below the
	saturated-ice phase \citep{hiroi_specific_2003,higashinaka_kagome_2004}.
Remarkably, within the experimental error, the relaxation times for decreasing and increasing field
    are of the same order of magnitude.
\\
%It should be noted that the determination of relaxation times
%    for field jumps upwords between 0.6~T and 1.0~T is difficult because the quench steps are only 0.1~T and, therefore,
%    the effects are masked by the minimum in the magnetostriction curve (see Fig.~\ref{fig:magnetostriction_trans_long}).
%    For this reason, the course there is marked by a dashed line.

%\Cref{fig:relaxationtimedto} also includes the data at those fields where long relaxation times were
%	measured. %, compare with \cref{fig:relaxationdto}.
%The relaxation times increase strongly with decreasing temperature for
%	the fields between \SI{0.5}{} to \SI{1.0}{\tesla}.

%\paragraph{Phase diagram}

\begin{figure}
	\centering
	\includegraphics[width=\linewidth]{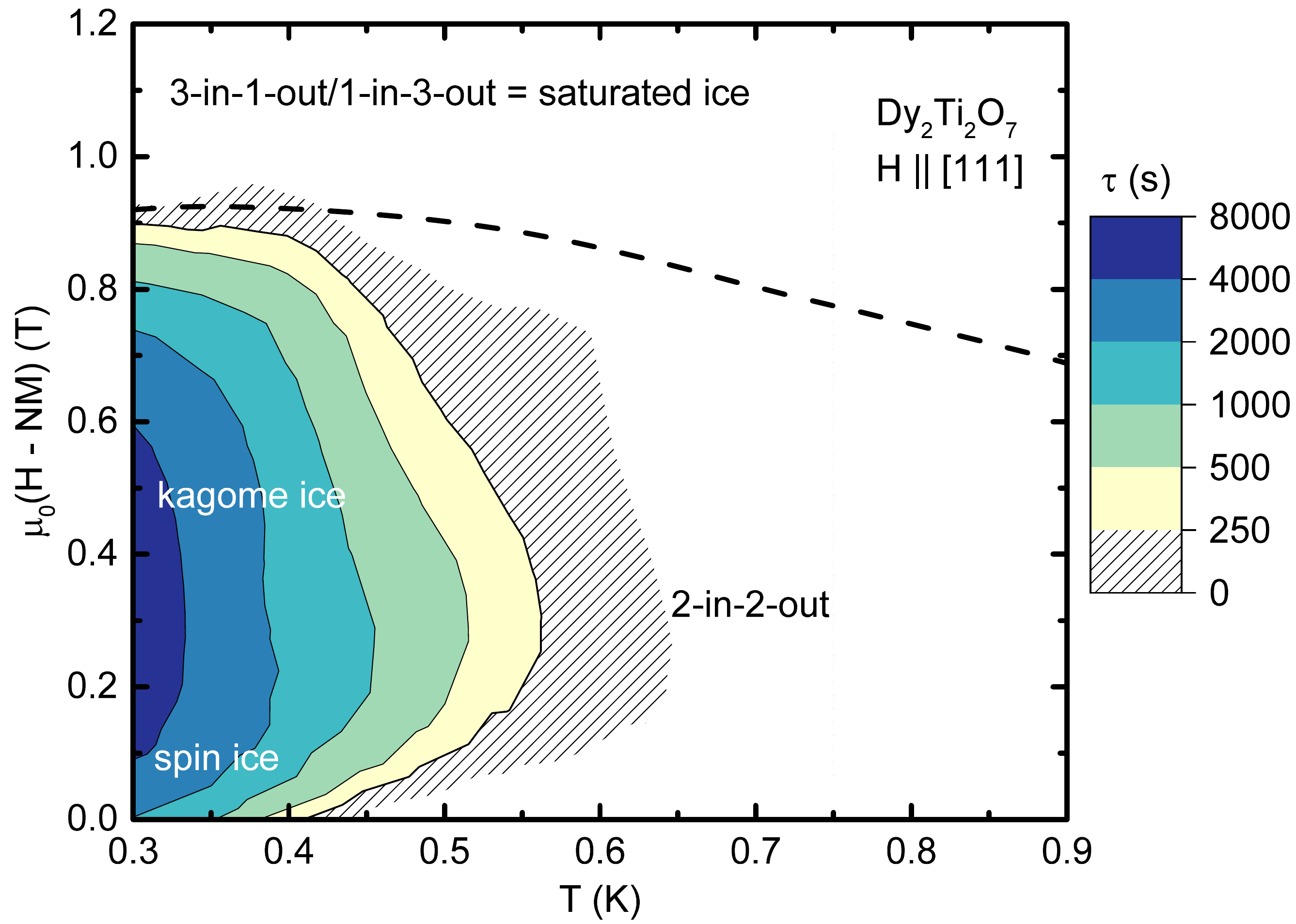}
	\caption{Contour plot of the relaxation time of \dto{} in the phase space of
	temperature and internal field with long relaxation times in blue
	and short relaxation fading to lighter colors.
	The relaxation times are taken from the quench experiments summarized
    in Fig.~\ref{fig:relaxationtimedto}.
	Note the exponential scale of the color scale.
	For comparison, transitions extracted from specific-heat and
	magnetization data \citep{hiroi_specific_2003} are included as dashed
	line.
	\label{fig:reltimecontour}}
\end{figure}

Finally, we summarize the relaxation times of our field-quench experiments
	in a phase diagram of temperature and internal field
	(corrected for demagnetization effects) in Fig.~\ref{fig:reltimecontour}. The internal field is calculated using
	$B_\mathrm{int} = \mu_0 (H - NM)$, where $\mu_0 H$
	is the applied field, $N$ the demagnetizing factor, and $M$ the known magnetization~\cite{sakakibara_observation_2003}.
In accordance with the known phase diagram of \dto{} extracted from specific-heat
    and magnetization data \citep{hiroi_specific_2003}, the lattice relaxation
    is only observed below the transition from the kagome to saturated-ice phase
    (dashed line in Fig.~\ref{fig:reltimecontour}).
The temperature-dependent relaxation time decreases with increasing temperature. The region of our measurable times (minimum 30~s)
    extends to \SI{0.6}{\kelvin}, the spin-freezing temperature~\citep{snyder_low-temperature_2004}.
At higher temperatures no relaxation is observed.
The longest relaxation times are found in the kagome-ice phase at low
	temperatures. The number of monopole-antimonopole pairs is particularly large in this phase and, therefore, in the kagome-ice
    phase long relaxation times can be expected.\\

To discuss the experimental facts, the field-quench experiments make the dynamics of thermally activated monopoles in the
    kagome phase of spin-ice compounds experimentally accessible. The magnetostriction
    is a highly sensitive probe, especially in the kagome-ice phase where magnetization changes could not be resolved.
    Our investigations on \dto{} evidence the dynamical behavior in spin ice
    as it has been modeled in theory with monopole ``3-in-1-out'' or ``1-in-3-out'' excitations from the ground-state
	configuration of ``2-in-2-out''~\citep{mostame_tunable_2014,castelnovo_thermal_2010}.
Specifically, a significant increase of the time scales of these dynamics at temperatures below \SI{0.6}{\kelvin}
    has been found. A number of publications deal with the dynamics of pyrochlore compounds, especially with \dto{}
    ~\citep{matsuhira_novel_2001, snyder_quantum-classical_2003,snyder_low-temperature_2004,
	matsuhira_spin_2011,yaraskavitch_spin_2012, takatsu_ac_2013,
	matsuhira_low_2000,snyder_how_2001,ehlers_dynamical_2003,
	ehlers_evidence_2004,quilliam_dynamics_2011,jackson_dynamic_2014,orendac_magnetocaloric_2007,
    erfanifam_intrinsic_2011,li_low-temperature_2015,pomaranski_absence_2013}.

%   Likewise, this increase has been observed in several different measurements on \dto{} such as ac
%	susceptibility \citep{matsuhira_novel_2001,
%	snyder_quantum-classical_2003,snyder_low-temperature_2004,
%	matsuhira_spin_2011,yaraskavitch_spin_2012, takatsu_ac_2013,
%	matsuhira_low_2000,snyder_how_2001,ehlers_dynamical_2003,
%	ehlers_evidence_2004,quilliam_dynamics_2011}, magnetization
%	\citep{matsuhira_spin_2011,paulsen_far-from-equilibrium_2014,
%	jackson_dynamic_2014}, magnetocaloric effect
%	\citep{orendac_magnetocaloric_2007}, ultrasound velocity
%	\citep{erfanifam_intrinsic_2011}, heat conductivity
%	\citep{li_low-temperature_2015}, and heat capacity
%	\citep{pomaranski_absence_2013}.

Many earlier investigations show dynamic effects below \SI{1}{\kelvin} on
	the same order of magnitude as the lattice relaxation and probably having
	the same origin. Our results confirm that the crystal lattice is active in the
    magnetic relaxation processes and directly reflects this relaxation.
Furthermore, since the relaxation times for increasing and decreasing field are almost identical,
    we experimentally prove that the formation and decay
    of monopole pairs follow the same time scales. Our measurements also show an increase of the
    relaxation times especially in the kagome-ice phase where we could follow the monopole
    dynamics over long time periods. This is in good agreement with theoretical predictions~\cite{mostame_tunable_2014}.
    The microscopic picture is that monopole-antimonopole pairs (``3-in-1-out'' or vice-versa
    configurations) on neighboring tetrahedra form stable
	bound pairs, that can neither annihilate nor move away from each other
	due to their mutual interaction. Therefore, the monopole mobility or spin-flip rate is reduced.
    Consequently, the probability of the annihilation of monopoles is
	suppressed and it takes a long time for this process to happen.
Additionally, monopole movement (spin flips) might be suppressed at defect sites
	of the lattice slowing down the intrinsic dynamics~\citep{bloxsom_thermal_2016}; this was suggested
    to explain the difference between the long-time thermal relaxation in specific-heat
    measurements~\citep{pomaranski_absence_2013}
    in comparison to other experiments~\citep{giblin_pauling_2018,henelius_refrustration_2016}.

In summary, we have shown dilatometric results of the lattice
	relaxation due to monopole dynamics in the classical spin ice compound \dto{}.
%	and \hto{}.
The lattice relaxation follows a stretched-exponential law with
	temperature- and field-dependent relaxation times $\tau$.
The analyzed field dependence of the relaxation, presented for the first time, illustrates the different character of
    the individual states. Extremely long relaxation times were observed in the kagome-ice. Our results fit very well
    to the behavior expected from theoretical considerations.
%for \dto{} and \SI{1.5}{\tesla} for \hto{}.
%We have observed similar relaxation processes also in the transverse magnetostriction of \dto{} (not shown).

%For \hto{}, similar features in the relaxation were found together with other anomalies as kinks where
%	the relaxation drastically changed its relaxation time (the data are published elsewhere).
%All these experimental facts agree well with the theoretical predictions of monopole properties
%    in spin-ice compounds, and show dilatometry as a suitable macroscopic method for the
%    investigation and understanding of dynamic effects in frustrated magnets.

\section{Acknowledgments}

We acknowledge support from HLD at HZDR, member of the European Magnetic Field
	Laboratory (EMFL).
This research has been supported by the DFG through SFB 1143
	(project-id 247310070) and the excellence cluster $ct.qmat$ (EXC 2147, project-id 39085490).
The work at the University of Warwick was supported by a grant from the EPSRC, UK (EP/M028771/1).
A portion of this work was performed at the NHMFL, which is supported by the NSF Cooperative Agreement No. DMR-1644779
    and the State of Florida.

\bibliography{dissertationnew}

\begin{thebibliography}{58}%
\makeatletter
\providecommand \@ifxundefined [1]{%
 \@ifx{#1\undefined}
}%
\providecommand \@ifnum [1]{%
 \ifnum #1\expandafter \@firstoftwo
 \else \expandafter \@secondoftwo
 \fi
}%
\providecommand \@ifx [1]{%
 \ifx #1\expandafter \@firstoftwo
 \else \expandafter \@secondoftwo
 \fi
}%
\providecommand \natexlab [1]{#1}%
\providecommand \enquote  [1]{``#1''}%
\providecommand \bibnamefont  [1]{#1}%
\providecommand \bibfnamefont [1]{#1}%
\providecommand \citenamefont [1]{#1}%
\providecommand \href@noop [0]{\@secondoftwo}%
\providecommand \href [0]{\begingroup \@sanitize@url \@href}%
\providecommand \@href[1]{\@@startlink{#1}\@@href}%
\providecommand \@@href[1]{\endgroup#1\@@endlink}%
\providecommand \@sanitize@url [0]{\catcode `\\12\catcode `\$12\catcode
  `\&12\catcode `\#12\catcode `\^12\catcode `\_12\catcode `\%12\relax}%
\providecommand \@@startlink[1]{}%
\providecommand \@@endlink[0]{}%
\providecommand \url  [0]{\begingroup\@sanitize@url \@url }%
\providecommand \@url [1]{\endgroup\@href {#1}{\urlprefix }}%
\providecommand \urlprefix  [0]{URL }%
\providecommand \Eprint [0]{\href }%
\providecommand \doibase [0]{http://dx.doi.org/}%
\providecommand \selectlanguage [0]{\@gobble}%
\providecommand \bibinfo  [0]{\@secondoftwo}%
\providecommand \bibfield  [0]{\@secondoftwo}%
\providecommand \translation [1]{[#1]}%
\providecommand \BibitemOpen [0]{}%
\providecommand \bibitemStop [0]{}%
\providecommand \bibitemNoStop [0]{.\EOS\space}%
\providecommand \EOS [0]{\spacefactor3000\relax}%
\providecommand \BibitemShut  [1]{\csname bibitem#1\endcsname}%
\let\auto@bib@innerbib\@empty
%</preamble>
\bibitem [{\citenamefont {Moessner}\ and\ \citenamefont
  {Ramirez}(2006)}]{moessner_pt_2006}%
  \BibitemOpen
  \bibfield  {author} {\bibinfo {author} {\bibfnamefont {R.}~\bibnamefont
  {Moessner}}\ and\ \bibinfo {author} {\bibfnamefont {A.}~\bibnamefont
  {Ramirez}},\ }\href@noop {} {\bibfield  {journal} {\bibinfo  {journal} {Phys.
  Today}\ }\textbf {\bibinfo {volume} {59(2)}},\ \bibinfo {pages} {24}
  (\bibinfo {year} {2006})}\BibitemShut {NoStop}%
\bibitem [{\citenamefont {Balents}(2010)}]{balents_spin_2010}%
  \BibitemOpen
  \bibfield  {author} {\bibinfo {author} {\bibfnamefont {L.}~\bibnamefont
  {Balents}},\ }\href {http://www.nature.com/doifinder/10.1038/nature08917}
  {\bibfield  {journal} {\bibinfo  {journal} {Nature}\ }\textbf {\bibinfo
  {volume} {464}},\ \bibinfo {pages} {199} (\bibinfo {year}
  {2010})}\BibitemShut {NoStop}%
\bibitem [{\citenamefont {Castelnovo}\ \emph {et~al.}(2008)\citenamefont
  {Castelnovo}, \citenamefont {Moessner},\ and\ \citenamefont
  {Sondhi}}]{castelnovo_magnetic_2008}%
  \BibitemOpen
  \bibfield  {author} {\bibinfo {author} {\bibfnamefont {C.}~\bibnamefont
  {Castelnovo}}, \bibinfo {author} {\bibfnamefont {R.}~\bibnamefont
  {Moessner}}, \ and\ \bibinfo {author} {\bibfnamefont {S.~L.}\ \bibnamefont
  {Sondhi}},\ }\href {http://www.nature.com/doifinder/10.1038/nature06433}
  {\bibfield  {journal} {\bibinfo  {journal} {Nature}\ }\textbf {\bibinfo
  {volume} {451}},\ \bibinfo {pages} {42} (\bibinfo {year} {2008})}\BibitemShut
  {NoStop}%
\bibitem [{\citenamefont {Slobinsky}\ \emph {et~al.}(2010)\citenamefont
  {Slobinsky}, \citenamefont {Castelnovo}, \citenamefont {Borzi}, \citenamefont
  {Gibbs}, \citenamefont {Mackenzie}, \citenamefont {Moessner},\ and\
  \citenamefont {Grigera}}]{slobinsky_unconventional_2010}%
  \BibitemOpen
  \bibfield  {author} {\bibinfo {author} {\bibfnamefont {D.}~\bibnamefont
  {Slobinsky}}, \bibinfo {author} {\bibfnamefont {C.}~\bibnamefont
  {Castelnovo}}, \bibinfo {author} {\bibfnamefont {R.~A.}\ \bibnamefont
  {Borzi}}, \bibinfo {author} {\bibfnamefont {A.~S.}\ \bibnamefont {Gibbs}},
  \bibinfo {author} {\bibfnamefont {A.~P.}\ \bibnamefont {Mackenzie}}, \bibinfo
  {author} {\bibfnamefont {R.}~\bibnamefont {Moessner}}, \ and\ \bibinfo
  {author} {\bibfnamefont {S.~A.}\ \bibnamefont {Grigera}},\ }\href
  {https://link.aps.org/doi/10.1103/PhysRevLett.105.267205} {\bibfield
  {journal} {\bibinfo  {journal} {Phys. Rev. Lett.}\ }\textbf {\bibinfo
  {volume} {105}},\ \bibinfo {pages} {267205} (\bibinfo {year}
  {2010})}\BibitemShut {NoStop}%
\bibitem [{\citenamefont {Erfanifam}\ \emph {et~al.}(2011)\citenamefont
  {Erfanifam}, \citenamefont {Zherlitsyn}, \citenamefont {Wosnitza},
  \citenamefont {Moessner}, \citenamefont {Petrenko}, \citenamefont
  {Balakrishnan},\ and\ \citenamefont {Zvyagin}}]{erfanifam_intrinsic_2011}%
  \BibitemOpen
  \bibfield  {author} {\bibinfo {author} {\bibfnamefont {S.}~\bibnamefont
  {Erfanifam}}, \bibinfo {author} {\bibfnamefont {S.}~\bibnamefont
  {Zherlitsyn}}, \bibinfo {author} {\bibfnamefont {J.}~\bibnamefont
  {Wosnitza}}, \bibinfo {author} {\bibfnamefont {R.}~\bibnamefont {Moessner}},
  \bibinfo {author} {\bibfnamefont {O.~A.}\ \bibnamefont {Petrenko}}, \bibinfo
  {author} {\bibfnamefont {G.}~\bibnamefont {Balakrishnan}}, \ and\ \bibinfo
  {author} {\bibfnamefont {A.~A.}\ \bibnamefont {Zvyagin}},\ }\href
  {http://link.aps.org/doi/10.1103/PhysRevB.84.220404} {\bibfield  {journal}
  {\bibinfo  {journal} {Phys. Rev. B}\ }\textbf {\bibinfo {volume} {84}},\
  \bibinfo {pages} {220404(R)} (\bibinfo {year} {2011})}\BibitemShut {NoStop}%
\bibitem [{\citenamefont {Erfanifam}\ \emph {et~al.}(2014)\citenamefont
  {Erfanifam}, \citenamefont {Zherlitsyn}, \citenamefont {Yasin}, \citenamefont
  {Skourski}, \citenamefont {Wosnitza}, \citenamefont {Zvyagin}, \citenamefont
  {McClarty}, \citenamefont {Moessner}, \citenamefont {Balakrishnan},\ and\
  \citenamefont {Petrenko}}]{erfanifam_ultrasonic_2014}%
  \BibitemOpen
  \bibfield  {author} {\bibinfo {author} {\bibfnamefont {S.}~\bibnamefont
  {Erfanifam}}, \bibinfo {author} {\bibfnamefont {S.}~\bibnamefont
  {Zherlitsyn}}, \bibinfo {author} {\bibfnamefont {S.}~\bibnamefont {Yasin}},
  \bibinfo {author} {\bibfnamefont {Y.}~\bibnamefont {Skourski}}, \bibinfo
  {author} {\bibfnamefont {J.}~\bibnamefont {Wosnitza}}, \bibinfo {author}
  {\bibfnamefont {A.~A.}\ \bibnamefont {Zvyagin}}, \bibinfo {author}
  {\bibfnamefont {P.}~\bibnamefont {McClarty}}, \bibinfo {author}
  {\bibfnamefont {R.}~\bibnamefont {Moessner}}, \bibinfo {author}
  {\bibfnamefont {G.}~\bibnamefont {Balakrishnan}}, \ and\ \bibinfo {author}
  {\bibfnamefont {O.~A.}\ \bibnamefont {Petrenko}},\ }\href
  {http://link.aps.org/doi/10.1103/PhysRevB.90.064409} {\bibfield  {journal}
  {\bibinfo  {journal} {Phys. Rev. B}\ }\textbf {\bibinfo {volume} {90}},\
  \bibinfo {pages} {064409} (\bibinfo {year} {2014})}\BibitemShut {NoStop}%
\bibitem [{\citenamefont {Borzi}\ \emph {et~al.}(2016)\citenamefont {Borzi},
  \citenamefont {Gomez~Albarracin}, \citenamefont {Rosales}, \citenamefont
  {Rossini}, \citenamefont {Steppke}, \citenamefont {Prabhakaran},
  \citenamefont {Mackenzie}, \citenamefont {Cabra},\ and\ \citenamefont
  {Grigera}}]{borzi_nc_2016}%
  \BibitemOpen
  \bibfield  {author} {\bibinfo {author} {\bibfnamefont {R.}~\bibnamefont
  {Borzi}}, \bibinfo {author} {\bibfnamefont {F.}~\bibnamefont
  {Gomez~Albarracin}}, \bibinfo {author} {\bibfnamefont {H.}~\bibnamefont
  {Rosales}}, \bibinfo {author} {\bibfnamefont {G.}~\bibnamefont {Rossini}},
  \bibinfo {author} {\bibfnamefont {A.}~\bibnamefont {Steppke}}, \bibinfo
  {author} {\bibfnamefont {D.}~\bibnamefont {Prabhakaran}}, \bibinfo {author}
  {\bibfnamefont {A.}~\bibnamefont {Mackenzie}}, \bibinfo {author}
  {\bibfnamefont {D.}~\bibnamefont {Cabra}}, \ and\ \bibinfo {author}
  {\bibfnamefont {S.}~\bibnamefont {Grigera}},\ }\href@noop {} {\bibfield
  {journal} {\bibinfo  {journal} {Nat. Commun.}\ }\textbf {\bibinfo {volume}
  {7}},\ \bibinfo {pages} {12592} (\bibinfo {year} {2016})}\BibitemShut
  {NoStop}%
\bibitem [{\citenamefont {Kaiser}\ \emph {et~al.}(2015)\citenamefont {Kaiser},
  \citenamefont {Bramwell}, \citenamefont {Holdsworth},\ and\ \citenamefont
  {Moessner}}]{kaiser_2015}%
  \BibitemOpen
  \bibfield  {author} {\bibinfo {author} {\bibfnamefont {V.}~\bibnamefont
  {Kaiser}}, \bibinfo {author} {\bibfnamefont {S.}~\bibnamefont {Bramwell}},
  \bibinfo {author} {\bibfnamefont {P.}~\bibnamefont {Holdsworth}}, \ and\
  \bibinfo {author} {\bibfnamefont {R.}~\bibnamefont {Moessner}},\ }\href@noop
  {} {\bibfield  {journal} {\bibinfo  {journal} {Phys. Rev. Lett.}\ }\textbf
  {\bibinfo {volume} {115}},\ \bibinfo {pages} {037201} (\bibinfo {year}
  {2015})}\BibitemShut {NoStop}%
\bibitem [{\citenamefont {Paulsen}\ \emph {et~al.}(2019)\citenamefont
  {Paulsen}, \citenamefont {Giblin}, \citenamefont {Lhotel}, \citenamefont
  {Prabhakaran}, \citenamefont {Matsuhira}, \citenamefont {Balakrishnan},\ and\
  \citenamefont {Bramwell}}]{paulsen_nc_2019}%
  \BibitemOpen
  \bibfield  {author} {\bibinfo {author} {\bibfnamefont {C.}~\bibnamefont
  {Paulsen}}, \bibinfo {author} {\bibfnamefont {S.}~\bibnamefont {Giblin}},
  \bibinfo {author} {\bibfnamefont {E.}~\bibnamefont {Lhotel}}, \bibinfo
  {author} {\bibfnamefont {D.}~\bibnamefont {Prabhakaran}}, \bibinfo {author}
  {\bibfnamefont {K.}~\bibnamefont {Matsuhira}}, \bibinfo {author}
  {\bibfnamefont {G.}~\bibnamefont {Balakrishnan}}, \ and\ \bibinfo {author}
  {\bibfnamefont {S.}~\bibnamefont {Bramwell}},\ }\href@noop {} {\bibfield
  {journal} {\bibinfo  {journal} {Nat. Commun.}\ }\textbf {\bibinfo {volume}
  {10}},\ \bibinfo {pages} {1509} (\bibinfo {year} {2019})}\BibitemShut
  {NoStop}%
\bibitem [{\citenamefont {Paulsen}\ \emph {et~al.}(2016)\citenamefont
  {Paulsen}, \citenamefont {Giblin}, \citenamefont {Lhotel}, \citenamefont
  {Canals}, \citenamefont {Prabhakaran}, \citenamefont {Matsuhira},\ and\
  \citenamefont {Bramwell}}]{paulsen_np_2016}%
  \BibitemOpen
  \bibfield  {author} {\bibinfo {author} {\bibfnamefont {C.}~\bibnamefont
  {Paulsen}}, \bibinfo {author} {\bibfnamefont {S.}~\bibnamefont {Giblin}},
  \bibinfo {author} {\bibfnamefont {E.}~\bibnamefont {Lhotel}}, \bibinfo
  {author} {\bibfnamefont {B.}~\bibnamefont {Canals}}, \bibinfo {author}
  {\bibfnamefont {D.}~\bibnamefont {Prabhakaran}}, \bibinfo {author}
  {\bibfnamefont {K.}~\bibnamefont {Matsuhira}}, \ and\ \bibinfo {author}
  {\bibfnamefont {S.}~\bibnamefont {Bramwell}},\ }\href@noop {} {\bibfield
  {journal} {\bibinfo  {journal} {Nat. Phys.}\ }\textbf {\bibinfo {volume}
  {12}},\ \bibinfo {pages} {661} (\bibinfo {year} {2016})}\BibitemShut
  {NoStop}%
\bibitem [{\citenamefont {Paulsen}\ \emph {et~al.}(2014)\citenamefont
  {Paulsen}, \citenamefont {Jackson}, \citenamefont {Lhotel}, \citenamefont
  {Canals}, \citenamefont {Prabhakaran}, \citenamefont {Matsuhira},
  \citenamefont {Giblin},\ and\ \citenamefont
  {Bramwell}}]{paulsen_far-from-equilibrium_2014}%
  \BibitemOpen
  \bibfield  {author} {\bibinfo {author} {\bibfnamefont {C.}~\bibnamefont
  {Paulsen}}, \bibinfo {author} {\bibfnamefont {M.~J.}\ \bibnamefont
  {Jackson}}, \bibinfo {author} {\bibfnamefont {E.}~\bibnamefont {Lhotel}},
  \bibinfo {author} {\bibfnamefont {B.}~\bibnamefont {Canals}}, \bibinfo
  {author} {\bibfnamefont {D.}~\bibnamefont {Prabhakaran}}, \bibinfo {author}
  {\bibfnamefont {K.}~\bibnamefont {Matsuhira}}, \bibinfo {author}
  {\bibfnamefont {S.~R.}\ \bibnamefont {Giblin}}, \ and\ \bibinfo {author}
  {\bibfnamefont {S.~T.}\ \bibnamefont {Bramwell}},\ }\href
  {http://www.nature.com/doifinder/10.1038/nphys2847} {\bibfield  {journal}
  {\bibinfo  {journal} {Nat. Phys.}\ }\textbf {\bibinfo {volume} {10}},\
  \bibinfo {pages} {135} (\bibinfo {year} {2014})}\BibitemShut {NoStop}%
\bibitem [{\citenamefont {Yaraskavitch}\ \emph {et~al.}(2012)\citenamefont
  {Yaraskavitch}, \citenamefont {Revell}, \citenamefont {Meng}, \citenamefont
  {Ross}, \citenamefont {Noad}, \citenamefont {Dabkowska}, \citenamefont
  {Gaulin},\ and\ \citenamefont {Kycia}}]{yaraskavitch_spin_2012}%
  \BibitemOpen
  \bibfield  {author} {\bibinfo {author} {\bibfnamefont {L.~R.}\ \bibnamefont
  {Yaraskavitch}}, \bibinfo {author} {\bibfnamefont {H.~M.}\ \bibnamefont
  {Revell}}, \bibinfo {author} {\bibfnamefont {S.}~\bibnamefont {Meng}},
  \bibinfo {author} {\bibfnamefont {K.~A.}\ \bibnamefont {Ross}}, \bibinfo
  {author} {\bibfnamefont {H.~M.~L.}\ \bibnamefont {Noad}}, \bibinfo {author}
  {\bibfnamefont {H.~A.}\ \bibnamefont {Dabkowska}}, \bibinfo {author}
  {\bibfnamefont {B.~D.}\ \bibnamefont {Gaulin}}, \ and\ \bibinfo {author}
  {\bibfnamefont {J.~B.}\ \bibnamefont {Kycia}},\ }\href
  {http://link.aps.org/doi/10.1103/PhysRevB.85.020410} {\bibfield  {journal}
  {\bibinfo  {journal} {Phys. Rev. B}\ }\textbf {\bibinfo {volume} {85}},\
  \bibinfo {pages} {020410(R)} (\bibinfo {year} {2012})}\BibitemShut {NoStop}%
\bibitem [{\citenamefont {Giblin}\ \emph
  {et~al.}(2018{\natexlab{a}})\citenamefont {Giblin}, \citenamefont
  {Twengstr\"om}, \citenamefont {Bovo}, \citenamefont {Ruminy}, \citenamefont
  {Bartkowiak}, \citenamefont {Manuel}, \citenamefont {Andresen}, \citenamefont
  {Prabhakaran}, \citenamefont {Balakrishnan}, \citenamefont {Pomjakushina},
  \citenamefont {Paulsen}, \citenamefont {Lhotel}, \citenamefont {Keller},
  \citenamefont {Frontzek}, \citenamefont {Capelli}, \citenamefont {Zaharko},
  \citenamefont {McClarty}, \citenamefont {Bramwell}, \citenamefont
  {Henelius},\ and\ \citenamefont {Fennell}}]{giblin_prl_2018}%
  \BibitemOpen
  \bibfield  {author} {\bibinfo {author} {\bibfnamefont {S.}~\bibnamefont
  {Giblin}}, \bibinfo {author} {\bibfnamefont {M.}~\bibnamefont
  {Twengstr\"om}}, \bibinfo {author} {\bibfnamefont {L.}~\bibnamefont {Bovo}},
  \bibinfo {author} {\bibfnamefont {M.}~\bibnamefont {Ruminy}}, \bibinfo
  {author} {\bibfnamefont {M.}~\bibnamefont {Bartkowiak}}, \bibinfo {author}
  {\bibfnamefont {P.}~\bibnamefont {Manuel}}, \bibinfo {author} {\bibfnamefont
  {J.}~\bibnamefont {Andresen}}, \bibinfo {author} {\bibfnamefont
  {D.}~\bibnamefont {Prabhakaran}}, \bibinfo {author} {\bibfnamefont
  {G.}~\bibnamefont {Balakrishnan}}, \bibinfo {author} {\bibfnamefont
  {E.}~\bibnamefont {Pomjakushina}}, \bibinfo {author} {\bibfnamefont
  {C.}~\bibnamefont {Paulsen}}, \bibinfo {author} {\bibfnamefont
  {E.}~\bibnamefont {Lhotel}}, \bibinfo {author} {\bibfnamefont
  {L.}~\bibnamefont {Keller}}, \bibinfo {author} {\bibfnamefont
  {M.}~\bibnamefont {Frontzek}}, \bibinfo {author} {\bibfnamefont
  {S.}~\bibnamefont {Capelli}}, \bibinfo {author} {\bibfnamefont
  {O.}~\bibnamefont {Zaharko}}, \bibinfo {author} {\bibfnamefont
  {P.}~\bibnamefont {McClarty}}, \bibinfo {author} {\bibfnamefont
  {S.}~\bibnamefont {Bramwell}}, \bibinfo {author} {\bibfnamefont
  {P.}~\bibnamefont {Henelius}}, \ and\ \bibinfo {author} {\bibfnamefont
  {T.}~\bibnamefont {Fennell}},\ }\href@noop {} {\bibfield  {journal} {\bibinfo
   {journal} {Phys. Rev. Lett.}\ }\textbf {\bibinfo {volume} {121}},\ \bibinfo
  {pages} {067202} (\bibinfo {year} {2018}{\natexlab{a}})}\BibitemShut
  {NoStop}%
\bibitem [{\citenamefont {Rau}\ and\ \citenamefont
  {Gingras}(2015)}]{rau_magnitude_2015}%
  \BibitemOpen
  \bibfield  {author} {\bibinfo {author} {\bibfnamefont {J.~G.}\ \bibnamefont
  {Rau}}\ and\ \bibinfo {author} {\bibfnamefont {M.~J.~P.}\ \bibnamefont
  {Gingras}},\ }\href {http://link.aps.org/doi/10.1103/PhysRevB.92.144417}
  {\bibfield  {journal} {\bibinfo  {journal} {Phys. Rev. B}\ }\textbf {\bibinfo
  {volume} {92}},\ \bibinfo {pages} {144417} (\bibinfo {year}
  {2015})}\BibitemShut {NoStop}%
\bibitem [{\citenamefont {den Hertog}\ and\ \citenamefont
  {Gingras}(2000)}]{den_hertog_dipolar_2000}%
  \BibitemOpen
  \bibfield  {author} {\bibinfo {author} {\bibfnamefont {B.~C.}\ \bibnamefont
  {den Hertog}}\ and\ \bibinfo {author} {\bibfnamefont {M.~J.~P.}\ \bibnamefont
  {Gingras}},\ }\href {http://link.aps.org/doi/10.1103/PhysRevLett.84.3430}
  {\bibfield  {journal} {\bibinfo  {journal} {Phys. Rev. Lett.}\ }\textbf
  {\bibinfo {volume} {84}},\ \bibinfo {pages} {3430} (\bibinfo {year}
  {2000})}\BibitemShut {NoStop}%
\bibitem [{\citenamefont {Bramwell}(2001)}]{bramwell_spin_2001}%
  \BibitemOpen
  \bibfield  {author} {\bibinfo {author} {\bibfnamefont {S.~T.}\ \bibnamefont
  {Bramwell}},\ }\href
  {http://www.sciencemag.org/cgi/doi/10.1126/science.1064761} {\bibfield
  {journal} {\bibinfo  {journal} {Science}\ }\textbf {\bibinfo {volume}
  {294}},\ \bibinfo {pages} {1495} (\bibinfo {year} {2001})}\BibitemShut
  {NoStop}%
\bibitem [{\citenamefont {Matsuhira}\ \emph {et~al.}(2001)\citenamefont
  {Matsuhira}, \citenamefont {Hinatsu},\ and\ \citenamefont
  {Sakakibara}}]{matsuhira_novel_2001}%
  \BibitemOpen
  \bibfield  {author} {\bibinfo {author} {\bibfnamefont {K.}~\bibnamefont
  {Matsuhira}}, \bibinfo {author} {\bibfnamefont {Y.}~\bibnamefont {Hinatsu}},
  \ and\ \bibinfo {author} {\bibfnamefont {T.}~\bibnamefont {Sakakibara}},\
  }\href
  {http://stacks.iop.org/0953-8984/13/i=31/a=101?key=crossref.6bca9132c3d363e56bacae695eceb8fd}
  {\bibfield  {journal} {\bibinfo  {journal} {J. Phys.: Condens. Mat.}\
  }\textbf {\bibinfo {volume} {13}},\ \bibinfo {pages} {L737} (\bibinfo {year}
  {2001})}\BibitemShut {NoStop}%
\bibitem [{\citenamefont {Snyder}\ \emph {et~al.}(2003)\citenamefont {Snyder},
  \citenamefont {Ueland}, \citenamefont {Slusky}, \citenamefont {Karunadasa},
  \citenamefont {Cava}, \citenamefont {Mizel},\ and\ \citenamefont
  {Schiffer}}]{snyder_quantum-classical_2003}%
  \BibitemOpen
  \bibfield  {author} {\bibinfo {author} {\bibfnamefont {J.}~\bibnamefont
  {Snyder}}, \bibinfo {author} {\bibfnamefont {B.~G.}\ \bibnamefont {Ueland}},
  \bibinfo {author} {\bibfnamefont {J.~S.}\ \bibnamefont {Slusky}}, \bibinfo
  {author} {\bibfnamefont {H.}~\bibnamefont {Karunadasa}}, \bibinfo {author}
  {\bibfnamefont {R.~J.}\ \bibnamefont {Cava}}, \bibinfo {author}
  {\bibfnamefont {A.}~\bibnamefont {Mizel}}, \ and\ \bibinfo {author}
  {\bibfnamefont {P.}~\bibnamefont {Schiffer}},\ }\href
  {http://link.aps.org/doi/10.1103/PhysRevLett.91.107201} {\bibfield  {journal}
  {\bibinfo  {journal} {Phys. Rev. Lett.}\ }\textbf {\bibinfo {volume} {91}},\
  \bibinfo {pages} {107201} (\bibinfo {year} {2003})}\BibitemShut {NoStop}%
\bibitem [{\citenamefont {Snyder}\ \emph {et~al.}(2004)\citenamefont {Snyder},
  \citenamefont {Ueland}, \citenamefont {Slusky}, \citenamefont {Karunadasa},
  \citenamefont {Cava},\ and\ \citenamefont
  {Schiffer}}]{snyder_low-temperature_2004}%
  \BibitemOpen
  \bibfield  {author} {\bibinfo {author} {\bibfnamefont {J.}~\bibnamefont
  {Snyder}}, \bibinfo {author} {\bibfnamefont {B.~G.}\ \bibnamefont {Ueland}},
  \bibinfo {author} {\bibfnamefont {J.~S.}\ \bibnamefont {Slusky}}, \bibinfo
  {author} {\bibfnamefont {H.}~\bibnamefont {Karunadasa}}, \bibinfo {author}
  {\bibfnamefont {R.~J.}\ \bibnamefont {Cava}}, \ and\ \bibinfo {author}
  {\bibfnamefont {P.}~\bibnamefont {Schiffer}},\ }\href
  {http://link.aps.org/doi/10.1103/PhysRevB.69.064414} {\bibfield  {journal}
  {\bibinfo  {journal} {Phys. Rev. B}\ }\textbf {\bibinfo {volume} {69}},\
  \bibinfo {pages} {064414} (\bibinfo {year} {2004})}\BibitemShut {NoStop}%
\bibitem [{\citenamefont {Matsuhira}\ \emph {et~al.}(2011)\citenamefont
  {Matsuhira}, \citenamefont {Paulsen}, \citenamefont {Lhotel}, \citenamefont
  {Sekine}, \citenamefont {Hiroi},\ and\ \citenamefont
  {Takagi}}]{matsuhira_spin_2011}%
  \BibitemOpen
  \bibfield  {author} {\bibinfo {author} {\bibfnamefont {K.}~\bibnamefont
  {Matsuhira}}, \bibinfo {author} {\bibfnamefont {C.}~\bibnamefont {Paulsen}},
  \bibinfo {author} {\bibfnamefont {E.}~\bibnamefont {Lhotel}}, \bibinfo
  {author} {\bibfnamefont {C.}~\bibnamefont {Sekine}}, \bibinfo {author}
  {\bibfnamefont {Z.}~\bibnamefont {Hiroi}}, \ and\ \bibinfo {author}
  {\bibfnamefont {S.}~\bibnamefont {Takagi}},\ }\href
  {http://journals.jps.jp/doi/abs/10.1143/JPSJ.80.123711} {\bibfield  {journal}
  {\bibinfo  {journal} {J. Phys. Soc. Jpn.}\ }\textbf {\bibinfo {volume}
  {80}},\ \bibinfo {pages} {123711} (\bibinfo {year} {2011})}\BibitemShut
  {NoStop}%
\bibitem [{\citenamefont {Matsuhira}\ \emph {et~al.}(2000)\citenamefont
  {Matsuhira}, \citenamefont {Hinatsu}, \citenamefont {Tenya},\ and\
  \citenamefont {Sakakibara}}]{matsuhira_low_2000}%
  \BibitemOpen
  \bibfield  {author} {\bibinfo {author} {\bibfnamefont {K.}~\bibnamefont
  {Matsuhira}}, \bibinfo {author} {\bibfnamefont {Y.}~\bibnamefont {Hinatsu}},
  \bibinfo {author} {\bibfnamefont {K.}~\bibnamefont {Tenya}}, \ and\ \bibinfo
  {author} {\bibfnamefont {T.}~\bibnamefont {Sakakibara}},\ }\href
  {http://stacks.iop.org/0953-8984/12/i=40/a=103?key=crossref.adc7c9d1d046c5325a02e88a46e193a8}
  {\bibfield  {journal} {\bibinfo  {journal} {J. Phys.: Condens. Mat.}\
  }\textbf {\bibinfo {volume} {12}},\ \bibinfo {pages} {L649} (\bibinfo {year}
  {2000})}\BibitemShut {NoStop}%
\bibitem [{\citenamefont {Snyder}\ \emph {et~al.}(2001)\citenamefont {Snyder},
  \citenamefont {Slusky}, \citenamefont {Cava},\ and\ \citenamefont
  {Schiffer}}]{snyder_how_2001}%
  \BibitemOpen
  \bibfield  {author} {\bibinfo {author} {\bibfnamefont {J.}~\bibnamefont
  {Snyder}}, \bibinfo {author} {\bibfnamefont {J.~S.}\ \bibnamefont {Slusky}},
  \bibinfo {author} {\bibfnamefont {R.~J.}\ \bibnamefont {Cava}}, \ and\
  \bibinfo {author} {\bibfnamefont {P.}~\bibnamefont {Schiffer}},\ }\href
  {http://www.nature.com/doifinder/10.1038/35092516} {\bibfield  {journal}
  {\bibinfo  {journal} {Nature}\ }\textbf {\bibinfo {volume} {413}},\ \bibinfo
  {pages} {48} (\bibinfo {year} {2001})}\BibitemShut {NoStop}%
\bibitem [{\citenamefont {Ehlers}\ \emph {et~al.}(2003)\citenamefont {Ehlers},
  \citenamefont {Cornelius}, \citenamefont {Orend{\'a}č}, \citenamefont
  {Kajnakov}, \citenamefont {Fennell}, \citenamefont {Bramwell},\ and\
  \citenamefont {Gardner}}]{ehlers_dynamical_2003}%
  \BibitemOpen
  \bibfield  {author} {\bibinfo {author} {\bibfnamefont {G.}~\bibnamefont
  {Ehlers}}, \bibinfo {author} {\bibfnamefont {A.~L.}\ \bibnamefont
  {Cornelius}}, \bibinfo {author} {\bibfnamefont {M.}~\bibnamefont
  {Orend{\'a}č}}, \bibinfo {author} {\bibfnamefont {M.}~\bibnamefont
  {Kajnakov}}, \bibinfo {author} {\bibfnamefont {T.}~\bibnamefont {Fennell}},
  \bibinfo {author} {\bibfnamefont {S.~T.}\ \bibnamefont {Bramwell}}, \ and\
  \bibinfo {author} {\bibfnamefont {J.~S.}\ \bibnamefont {Gardner}},\ }\href
  {http://stacks.iop.org/0953-8984/15/i=2/a=102?key=crossref.0967240c1fee62e5f6a98d94ab9cfaef}
  {\bibfield  {journal} {\bibinfo  {journal} {J. Phys.: Condens. Mat.}\
  }\textbf {\bibinfo {volume} {15}},\ \bibinfo {pages} {L9} (\bibinfo {year}
  {2003})}\BibitemShut {NoStop}%
\bibitem [{\citenamefont {Ehlers}\ \emph {et~al.}(2004)\citenamefont {Ehlers},
  \citenamefont {Cornelius}, \citenamefont {Fennell}, \citenamefont {Koza},
  \citenamefont {Bramwell},\ and\ \citenamefont
  {Gardner}}]{ehlers_evidence_2004}%
  \BibitemOpen
  \bibfield  {author} {\bibinfo {author} {\bibfnamefont {G.}~\bibnamefont
  {Ehlers}}, \bibinfo {author} {\bibfnamefont {A.~L.}\ \bibnamefont
  {Cornelius}}, \bibinfo {author} {\bibfnamefont {T.}~\bibnamefont {Fennell}},
  \bibinfo {author} {\bibfnamefont {M.}~\bibnamefont {Koza}}, \bibinfo {author}
  {\bibfnamefont {S.~T.}\ \bibnamefont {Bramwell}}, \ and\ \bibinfo {author}
  {\bibfnamefont {J.~S.}\ \bibnamefont {Gardner}},\ }\href
  {http://stacks.iop.org/0953-8984/16/i=11/a=010?key=crossref.9358f6b637ab5b58305de469cdf8f8f4}
  {\bibfield  {journal} {\bibinfo  {journal} {J. Phys.: Condens. Mat.}\
  }\textbf {\bibinfo {volume} {16}},\ \bibinfo {pages} {S635} (\bibinfo {year}
  {2004})}\BibitemShut {NoStop}%
\bibitem [{\citenamefont {Petrenko}\ \emph {et~al.}(2011)\citenamefont
  {Petrenko}, \citenamefont {Lees},\ and\ \citenamefont
  {Balakrishnan}}]{petrenko_mag_2011}%
  \BibitemOpen
  \bibfield  {author} {\bibinfo {author} {\bibfnamefont {O.~A.}\ \bibnamefont
  {Petrenko}}, \bibinfo {author} {\bibfnamefont {M.~R.}\ \bibnamefont {Lees}},
  \ and\ \bibinfo {author} {\bibfnamefont {G.}~\bibnamefont {Balakrishnan}},\
  }\href {https://doi.org/10.1088/0953-8984/23/16/164218} {\bibfield  {journal}
  {\bibinfo  {journal} {J. Phys.: Condens. Mat.}\ }\textbf {\bibinfo {volume}
  {23}},\ \bibinfo {pages} {164218} (\bibinfo {year} {2011})}\BibitemShut
  {NoStop}%
\bibitem [{\citenamefont {Quilliam}\ \emph {et~al.}(2011)\citenamefont
  {Quilliam}, \citenamefont {Yaraskavitch}, \citenamefont {Dabkowska},
  \citenamefont {Gaulin},\ and\ \citenamefont
  {Kycia}}]{quilliam_dynamics_2011}%
  \BibitemOpen
  \bibfield  {author} {\bibinfo {author} {\bibfnamefont {J.~A.}\ \bibnamefont
  {Quilliam}}, \bibinfo {author} {\bibfnamefont {L.~R.}\ \bibnamefont
  {Yaraskavitch}}, \bibinfo {author} {\bibfnamefont {H.~A.}\ \bibnamefont
  {Dabkowska}}, \bibinfo {author} {\bibfnamefont {B.~D.}\ \bibnamefont
  {Gaulin}}, \ and\ \bibinfo {author} {\bibfnamefont {J.~B.}\ \bibnamefont
  {Kycia}},\ }\href {http://link.aps.org/doi/10.1103/PhysRevB.83.094424}
  {\bibfield  {journal} {\bibinfo  {journal} {Phys. Rev. B}\ }\textbf {\bibinfo
  {volume} {83}},\ \bibinfo {pages} {094424} (\bibinfo {year}
  {2011})}\BibitemShut {NoStop}%
\bibitem [{\citenamefont {Jaubert}\ and\ \citenamefont
  {Holdsworth}(2009)}]{jaubert_magnetic_2009}%
  \BibitemOpen
  \bibfield  {author} {\bibinfo {author} {\bibfnamefont {L.~C.}\ \bibnamefont
  {Jaubert}}\ and\ \bibinfo {author} {\bibfnamefont {P.~C.~W.}\ \bibnamefont
  {Holdsworth}},\ }\href {http://www.nature.com/doifinder/10.1038/nphys1227}
  {\bibfield  {journal} {\bibinfo  {journal} {Nat. Phys.}\ }\textbf {\bibinfo
  {volume} {5}},\ \bibinfo {pages} {258} (\bibinfo {year} {2009})}\BibitemShut
  {NoStop}%
\bibitem [{\citenamefont {Jaubert}\ and\ \citenamefont
  {Holdsworth}(2011)}]{jaubert_magnetic_2011}%
  \BibitemOpen
  \bibfield  {author} {\bibinfo {author} {\bibfnamefont {L.~C.}\ \bibnamefont
  {Jaubert}}\ and\ \bibinfo {author} {\bibfnamefont {P.~C.~W.}\ \bibnamefont
  {Holdsworth}},\ }\href
  {http://stacks.iop.org/0953-8984/23/i=16/a=164222?key=crossref.e307b5b4d150d9ededdb668f0c7926f5}
  {\bibfield  {journal} {\bibinfo  {journal} {J. Phys.: Condens. Mat.}\
  }\textbf {\bibinfo {volume} {23}},\ \bibinfo {pages} {164222} (\bibinfo
  {year} {2011})}\BibitemShut {NoStop}%
\bibitem [{\citenamefont {Giblin}\ \emph {et~al.}(2011)\citenamefont {Giblin},
  \citenamefont {Bramwell}, \citenamefont {Holdsworth}, \citenamefont
  {Prabhakaran},\ and\ \citenamefont {Terry}}]{giblin_relax_2011}%
  \BibitemOpen
  \bibfield  {author} {\bibinfo {author} {\bibfnamefont {S.~R.}\ \bibnamefont
  {Giblin}}, \bibinfo {author} {\bibfnamefont {S.~T.}\ \bibnamefont
  {Bramwell}}, \bibinfo {author} {\bibfnamefont {P.~C.~W.}\ \bibnamefont
  {Holdsworth}}, \bibinfo {author} {\bibfnamefont {D.}~\bibnamefont
  {Prabhakaran}}, \ and\ \bibinfo {author} {\bibfnamefont {I.}~\bibnamefont
  {Terry}},\ }\href {http://www.nature.com/doifinder/10.1038/nphys1896}
  {\bibfield  {journal} {\bibinfo  {journal} {Nat. Phys.}\ }\textbf {\bibinfo
  {volume} {7}},\ \bibinfo {pages} {252} (\bibinfo {year} {2011})}\BibitemShut
  {NoStop}%
\bibitem [{\citenamefont {Castelnovo}\ \emph {et~al.}(2010)\citenamefont
  {Castelnovo}, \citenamefont {Moessner},\ and\ \citenamefont
  {Sondhi}}]{castelnovo_thermal_2010}%
  \BibitemOpen
  \bibfield  {author} {\bibinfo {author} {\bibfnamefont {C.}~\bibnamefont
  {Castelnovo}}, \bibinfo {author} {\bibfnamefont {R.}~\bibnamefont
  {Moessner}}, \ and\ \bibinfo {author} {\bibfnamefont {S.~L.}\ \bibnamefont
  {Sondhi}},\ }\href {https://link.aps.org/doi/10.1103/PhysRevLett.104.107201}
  {\bibfield  {journal} {\bibinfo  {journal} {Phys. Rev. Lett.}\ }\textbf
  {\bibinfo {volume} {104}},\ \bibinfo {pages} {107201} (\bibinfo {year}
  {2010})}\BibitemShut {NoStop}%
\bibitem [{\citenamefont {Mostame}\ \emph {et~al.}(2014)\citenamefont
  {Mostame}, \citenamefont {Castelnovo}, \citenamefont {Moessner},\ and\
  \citenamefont {Sondhi}}]{mostame_tunable_2014}%
  \BibitemOpen
  \bibfield  {author} {\bibinfo {author} {\bibfnamefont {S.}~\bibnamefont
  {Mostame}}, \bibinfo {author} {\bibfnamefont {C.}~\bibnamefont {Castelnovo}},
  \bibinfo {author} {\bibfnamefont {R.}~\bibnamefont {Moessner}}, \ and\
  \bibinfo {author} {\bibfnamefont {S.~L.}\ \bibnamefont {Sondhi}},\ }\href
  {http://www.pnas.org/cgi/doi/10.1073/pnas.1317631111} {\bibfield  {journal}
  {\bibinfo  {journal} {Proc. Natl. Acad. Sci. USA}\ }\textbf {\bibinfo
  {volume} {111}},\ \bibinfo {pages} {640} (\bibinfo {year}
  {2014})}\BibitemShut {NoStop}%
\bibitem [{\citenamefont {Yamashita}\ and\ \citenamefont
  {Ueda}(2000)}]{yamashita_spin-driven_2000}%
  \BibitemOpen
  \bibfield  {author} {\bibinfo {author} {\bibfnamefont {Y.}~\bibnamefont
  {Yamashita}}\ and\ \bibinfo {author} {\bibfnamefont {K.}~\bibnamefont
  {Ueda}},\ }\href {https://link.aps.org/doi/10.1103/PhysRevLett.85.4960}
  {\bibfield  {journal} {\bibinfo  {journal} {Phys. Rev. Lett.}\ }\textbf
  {\bibinfo {volume} {85}},\ \bibinfo {pages} {4960} (\bibinfo {year}
  {2000})}\BibitemShut {NoStop}%
\bibitem [{\citenamefont {Tchernyshyov}\ \emph
  {et~al.}(2002{\natexlab{a}})\citenamefont {Tchernyshyov}, \citenamefont
  {Moessner},\ and\ \citenamefont {Sondhi}}]{tchernyshyov_order_2002}%
  \BibitemOpen
  \bibfield  {author} {\bibinfo {author} {\bibfnamefont {O.}~\bibnamefont
  {Tchernyshyov}}, \bibinfo {author} {\bibfnamefont {R.}~\bibnamefont
  {Moessner}}, \ and\ \bibinfo {author} {\bibfnamefont {S.~L.}\ \bibnamefont
  {Sondhi}},\ }\href {https://link.aps.org/doi/10.1103/PhysRevLett.88.067203}
  {\bibfield  {journal} {\bibinfo  {journal} {Phys. Rev. Lett.}\ }\textbf
  {\bibinfo {volume} {88}},\ \bibinfo {pages} {067203} (\bibinfo {year}
  {2002}{\natexlab{a}})}\BibitemShut {NoStop}%
\bibitem [{\citenamefont {Tchernyshyov}\ \emph
  {et~al.}(2002{\natexlab{b}})\citenamefont {Tchernyshyov}, \citenamefont
  {Moessner},\ and\ \citenamefont {Sondhi}}]{tchernyshyov_spin-peierls_2002}%
  \BibitemOpen
  \bibfield  {author} {\bibinfo {author} {\bibfnamefont {O.}~\bibnamefont
  {Tchernyshyov}}, \bibinfo {author} {\bibfnamefont {R.}~\bibnamefont
  {Moessner}}, \ and\ \bibinfo {author} {\bibfnamefont {S.~L.}\ \bibnamefont
  {Sondhi}},\ }\href {https://link.aps.org/doi/10.1103/PhysRevB.66.064403}
  {\bibfield  {journal} {\bibinfo  {journal} {Phys. Rev. B}\ }\textbf {\bibinfo
  {volume} {66}},\ \bibinfo {pages} {064403} (\bibinfo {year}
  {2002}{\natexlab{b}})}\BibitemShut {NoStop}%
\bibitem [{\citenamefont {Richter}\ \emph {et~al.}(2004)\citenamefont
  {Richter}, \citenamefont {Derzhko},\ and\ \citenamefont
  {Schulenburg}}]{richter_magnetic-field_2004}%
  \BibitemOpen
  \bibfield  {author} {\bibinfo {author} {\bibfnamefont {J.}~\bibnamefont
  {Richter}}, \bibinfo {author} {\bibfnamefont {O.}~\bibnamefont {Derzhko}}, \
  and\ \bibinfo {author} {\bibfnamefont {J.}~\bibnamefont {Schulenburg}},\
  }\href {https://link.aps.org/doi/10.1103/PhysRevLett.93.107206} {\bibfield
  {journal} {\bibinfo  {journal} {Phys. Rev. Lett.}\ }\textbf {\bibinfo
  {volume} {93}},\ \bibinfo {pages} {107206} (\bibinfo {year}
  {2004})}\BibitemShut {NoStop}%
\bibitem [{\citenamefont {Jia}\ \emph {et~al.}(2005)\citenamefont {Jia},
  \citenamefont {Nam}, \citenamefont {Kim},\ and\ \citenamefont
  {Han}}]{jia_lattice-coupled_2005}%
  \BibitemOpen
  \bibfield  {author} {\bibinfo {author} {\bibfnamefont {C.}~\bibnamefont
  {Jia}}, \bibinfo {author} {\bibfnamefont {J.~H.}\ \bibnamefont {Nam}},
  \bibinfo {author} {\bibfnamefont {J.~S.}\ \bibnamefont {Kim}}, \ and\
  \bibinfo {author} {\bibfnamefont {J.~H.}\ \bibnamefont {Han}},\ }\href
  {https://link.aps.org/doi/10.1103/PhysRevB.71.212406} {\bibfield  {journal}
  {\bibinfo  {journal} {Phys. Rev. B}\ }\textbf {\bibinfo {volume} {71}},\
  \bibinfo {pages} {212406} (\bibinfo {year} {2005})}\BibitemShut {NoStop}%
\bibitem [{\citenamefont {Penc}\ \emph {et~al.}(2007)\citenamefont {Penc},
  \citenamefont {Shannon}, \citenamefont {Motome},\ and\ \citenamefont
  {Shiba}}]{penc_symmetry_2007}%
  \BibitemOpen
  \bibfield  {author} {\bibinfo {author} {\bibfnamefont {K.}~\bibnamefont
  {Penc}}, \bibinfo {author} {\bibfnamefont {N.}~\bibnamefont {Shannon}},
  \bibinfo {author} {\bibfnamefont {Y.}~\bibnamefont {Motome}}, \ and\ \bibinfo
  {author} {\bibfnamefont {H.}~\bibnamefont {Shiba}},\ }\href
  {http://stacks.iop.org/0953-8984/19/i=14/a=145267?key=crossref.687ea493522ba39b47fad2fc2a2d693e}
  {\bibfield  {journal} {\bibinfo  {journal} {J. Phys.: Condens. Mat.}\
  }\textbf {\bibinfo {volume} {19}},\ \bibinfo {pages} {145267} (\bibinfo
  {year} {2007})}\BibitemShut {NoStop}%
\bibitem [{\citenamefont {Terao}\ and\ \citenamefont
  {Honda}(2007)}]{terao_distortion_2007}%
  \BibitemOpen
  \bibfield  {author} {\bibinfo {author} {\bibfnamefont {K.}~\bibnamefont
  {Terao}}\ and\ \bibinfo {author} {\bibfnamefont {I.}~\bibnamefont {Honda}},\
  }\href
  {http://stacks.iop.org/0953-8984/19/i=14/a=145261?key=crossref.3bf4e42d9af33a1a69d8b65eb64d9cb4}
  {\bibfield  {journal} {\bibinfo  {journal} {J. Phys.: Condens. Mat.}\
  }\textbf {\bibinfo {volume} {19}},\ \bibinfo {pages} {145261} (\bibinfo
  {year} {2007})}\BibitemShut {NoStop}%
\bibitem [{\citenamefont {Saunders}\ and\ \citenamefont
  {Chalker}(2008)}]{saunders_structural_2008}%
  \BibitemOpen
  \bibfield  {author} {\bibinfo {author} {\bibfnamefont {T.~E.}\ \bibnamefont
  {Saunders}}\ and\ \bibinfo {author} {\bibfnamefont {J.~T.}\ \bibnamefont
  {Chalker}},\ }\href {https://link.aps.org/doi/10.1103/PhysRevB.77.214438}
  {\bibfield  {journal} {\bibinfo  {journal} {Phys. Rev. B}\ }\textbf {\bibinfo
  {volume} {77}},\ \bibinfo {pages} {214438} (\bibinfo {year}
  {2008})}\BibitemShut {NoStop}%
\bibitem [{\citenamefont {Curnoe}(2008)}]{curnoe_structural_2008}%
  \BibitemOpen
  \bibfield  {author} {\bibinfo {author} {\bibfnamefont {S.~H.}\ \bibnamefont
  {Curnoe}},\ }\href {http://link.aps.org/doi/10.1103/PhysRevB.78.094418}
  {\bibfield  {journal} {\bibinfo  {journal} {Phys. Rev. B}\ }\textbf {\bibinfo
  {volume} {78}},\ \bibinfo {pages} {094418} (\bibinfo {year}
  {2008})}\BibitemShut {NoStop}%
\bibitem [{\citenamefont {Aoyama}\ and\ \citenamefont
  {Kawamura}(2016)}]{aoyama_spin-lattice-coupled_2016}%
  \BibitemOpen
  \bibfield  {author} {\bibinfo {author} {\bibfnamefont {K.}~\bibnamefont
  {Aoyama}}\ and\ \bibinfo {author} {\bibfnamefont {H.}~\bibnamefont
  {Kawamura}},\ }\href
  {https://link.aps.org/doi/10.1103/PhysRevLett.116.257201} {\bibfield
  {journal} {\bibinfo  {journal} {Phys. Rev. Lett.}\ }\textbf {\bibinfo
  {volume} {116}},\ \bibinfo {pages} {257201} (\bibinfo {year}
  {2016})}\BibitemShut {NoStop}%
\bibitem [{\citenamefont {Mirebeau}\ and\ \citenamefont
  {Goncharenko}(2004)}]{mirebeau_spin_2004}%
  \BibitemOpen
  \bibfield  {author} {\bibinfo {author} {\bibfnamefont {I.}~\bibnamefont
  {Mirebeau}}\ and\ \bibinfo {author} {\bibfnamefont {I.}~\bibnamefont
  {Goncharenko}},\ }\href
  {http://stacks.iop.org/0953-8984/16/i=11/a=012?key=crossref.8412378d16047bd3ea1cdd4f94fd46e1}
  {\bibfield  {journal} {\bibinfo  {journal} {J. Phys.: Condens. Mat.}\
  }\textbf {\bibinfo {volume} {16}},\ \bibinfo {pages} {S653} (\bibinfo {year}
  {2004})}\BibitemShut {NoStop}%
\bibitem [{\citenamefont {Jackson}\ \emph {et~al.}(2014)\citenamefont
  {Jackson}, \citenamefont {Lhotel}, \citenamefont {Giblin}, \citenamefont
  {Bramwell}, \citenamefont {Prabhakaran}, \citenamefont {Matsuhira},
  \citenamefont {Hiroi}, \citenamefont {Yu},\ and\ \citenamefont
  {Paulsen}}]{jackson_dynamic_2014}%
  \BibitemOpen
  \bibfield  {author} {\bibinfo {author} {\bibfnamefont {M.~J.}\ \bibnamefont
  {Jackson}}, \bibinfo {author} {\bibfnamefont {E.}~\bibnamefont {Lhotel}},
  \bibinfo {author} {\bibfnamefont {S.~R.}\ \bibnamefont {Giblin}}, \bibinfo
  {author} {\bibfnamefont {S.~T.}\ \bibnamefont {Bramwell}}, \bibinfo {author}
  {\bibfnamefont {D.}~\bibnamefont {Prabhakaran}}, \bibinfo {author}
  {\bibfnamefont {K.}~\bibnamefont {Matsuhira}}, \bibinfo {author}
  {\bibfnamefont {Z.}~\bibnamefont {Hiroi}}, \bibinfo {author} {\bibfnamefont
  {Q.}~\bibnamefont {Yu}}, \ and\ \bibinfo {author} {\bibfnamefont
  {C.}~\bibnamefont {Paulsen}},\ }\href
  {http://link.aps.org/doi/10.1103/PhysRevB.90.064427} {\bibfield  {journal}
  {\bibinfo  {journal} {Phys. Rev. B}\ }\textbf {\bibinfo {volume} {90}},\
  \bibinfo {pages} {064427} (\bibinfo {year} {2014})}\BibitemShut {NoStop}%
\bibitem [{\citenamefont {Takatsu}\ \emph {et~al.}(2013)\citenamefont
  {Takatsu}, \citenamefont {Goto}, \citenamefont {Otsuka}, \citenamefont
  {Higashinaka}, \citenamefont {Matsubayashi}, \citenamefont {Uwatoko},\ and\
  \citenamefont {Kadowaki}}]{takatsu_ac_2013}%
  \BibitemOpen
  \bibfield  {author} {\bibinfo {author} {\bibfnamefont {H.}~\bibnamefont
  {Takatsu}}, \bibinfo {author} {\bibfnamefont {K.}~\bibnamefont {Goto}},
  \bibinfo {author} {\bibfnamefont {H.}~\bibnamefont {Otsuka}}, \bibinfo
  {author} {\bibfnamefont {R.}~\bibnamefont {Higashinaka}}, \bibinfo {author}
  {\bibfnamefont {K.}~\bibnamefont {Matsubayashi}}, \bibinfo {author}
  {\bibfnamefont {Y.}~\bibnamefont {Uwatoko}}, \ and\ \bibinfo {author}
  {\bibfnamefont {H.}~\bibnamefont {Kadowaki}},\ }\href
  {http://journals.jps.jp/doi/abs/10.7566/JPSJ.82.104710} {\bibfield  {journal}
  {\bibinfo  {journal} {J. Phys. Soc. Jpn.}\ }\textbf {\bibinfo {volume}
  {82}},\ \bibinfo {pages} {104710} (\bibinfo {year} {2013})}\BibitemShut
  {NoStop}%
\bibitem [{\citenamefont {Orendáč}\ \emph {et~al.}(2007)\citenamefont
  {Orendáč}, \citenamefont {Hanko}, \citenamefont {Čižmár}, \citenamefont
  {Orendáčová}, \citenamefont {Shirai},\ and\ \citenamefont
  {Bramwell}}]{orendac_magnetocaloric_2007}%
  \BibitemOpen
  \bibfield  {author} {\bibinfo {author} {\bibfnamefont {M.}~\bibnamefont
  {Orendáč}}, \bibinfo {author} {\bibfnamefont {J.}~\bibnamefont {Hanko}},
  \bibinfo {author} {\bibfnamefont {E.}~\bibnamefont {Čižmár}}, \bibinfo
  {author} {\bibfnamefont {A.}~\bibnamefont {Orendáčová}}, \bibinfo {author}
  {\bibfnamefont {M.}~\bibnamefont {Shirai}}, \ and\ \bibinfo {author}
  {\bibfnamefont {S.~T.}\ \bibnamefont {Bramwell}},\ }\href
  {http://link.aps.org/doi/10.1103/PhysRevB.75.104425} {\bibfield  {journal}
  {\bibinfo  {journal} {Phys. Rev. B}\ }\textbf {\bibinfo {volume} {75}},\
  \bibinfo {pages} {104425} (\bibinfo {year} {2007})}\BibitemShut {NoStop}%
\bibitem [{\citenamefont {Li}\ \emph {et~al.}(2015)\citenamefont {Li},
  \citenamefont {Zhao}, \citenamefont {Fan}, \citenamefont {Tong},
  \citenamefont {Zhang}, \citenamefont {Shi}, \citenamefont {Wu}, \citenamefont
  {Liu}, \citenamefont {Zhou}, \citenamefont {Zhao},\ and\ \citenamefont
  {Sun}}]{li_low-temperature_2015}%
  \BibitemOpen
  \bibfield  {author} {\bibinfo {author} {\bibfnamefont {S.~J.}\ \bibnamefont
  {Li}}, \bibinfo {author} {\bibfnamefont {Z.~Y.}\ \bibnamefont {Zhao}},
  \bibinfo {author} {\bibfnamefont {C.}~\bibnamefont {Fan}}, \bibinfo {author}
  {\bibfnamefont {B.}~\bibnamefont {Tong}}, \bibinfo {author} {\bibfnamefont
  {F.~B.}\ \bibnamefont {Zhang}}, \bibinfo {author} {\bibfnamefont
  {J.}~\bibnamefont {Shi}}, \bibinfo {author} {\bibfnamefont {J.~C.}\
  \bibnamefont {Wu}}, \bibinfo {author} {\bibfnamefont {X.~G.}\ \bibnamefont
  {Liu}}, \bibinfo {author} {\bibfnamefont {H.~D.}\ \bibnamefont {Zhou}},
  \bibinfo {author} {\bibfnamefont {X.}~\bibnamefont {Zhao}}, \ and\ \bibinfo
  {author} {\bibfnamefont {X.~F.}\ \bibnamefont {Sun}},\ }\href
  {http://link.aps.org/doi/10.1103/PhysRevB.92.094408} {\bibfield  {journal}
  {\bibinfo  {journal} {Phys. Rev. B}\ }\textbf {\bibinfo {volume} {92}},\
  \bibinfo {pages} {094408} (\bibinfo {year} {2015})}\BibitemShut {NoStop}%
\bibitem [{\citenamefont {Pomaranski}\ \emph {et~al.}(2013)\citenamefont
  {Pomaranski}, \citenamefont {Yaraskavitch}, \citenamefont {Meng},
  \citenamefont {Ross}, \citenamefont {Noad}, \citenamefont {Dabkowska},
  \citenamefont {Gaulin},\ and\ \citenamefont
  {Kycia}}]{pomaranski_absence_2013}%
  \BibitemOpen
  \bibfield  {author} {\bibinfo {author} {\bibfnamefont {D.}~\bibnamefont
  {Pomaranski}}, \bibinfo {author} {\bibfnamefont {L.~R.}\ \bibnamefont
  {Yaraskavitch}}, \bibinfo {author} {\bibfnamefont {S.}~\bibnamefont {Meng}},
  \bibinfo {author} {\bibfnamefont {K.~A.}\ \bibnamefont {Ross}}, \bibinfo
  {author} {\bibfnamefont {H.~M.~L.}\ \bibnamefont {Noad}}, \bibinfo {author}
  {\bibfnamefont {H.~A.}\ \bibnamefont {Dabkowska}}, \bibinfo {author}
  {\bibfnamefont {B.~D.}\ \bibnamefont {Gaulin}}, \ and\ \bibinfo {author}
  {\bibfnamefont {J.~B.}\ \bibnamefont {Kycia}},\ }\href
  {http://www.nature.com/doifinder/10.1038/nphys2591} {\bibfield  {journal}
  {\bibinfo  {journal} {Nat. Phys.}\ }\textbf {\bibinfo {volume} {9}},\
  \bibinfo {pages} {353} (\bibinfo {year} {2013})}\BibitemShut {NoStop}%
\bibitem [{\citenamefont {Balakrishnan}\ \emph {et~al.}(1998)\citenamefont
  {Balakrishnan}, \citenamefont {Petrenko}, \citenamefont {Lees},\ and\
  \citenamefont {Paul}}]{balakrishnan_single_1998}%
  \BibitemOpen
  \bibfield  {author} {\bibinfo {author} {\bibfnamefont {G.}~\bibnamefont
  {Balakrishnan}}, \bibinfo {author} {\bibfnamefont {O.~A.}\ \bibnamefont
  {Petrenko}}, \bibinfo {author} {\bibfnamefont {M.~R.}\ \bibnamefont {Lees}},
  \ and\ \bibinfo {author} {\bibfnamefont {D.~M.}\ \bibnamefont {Paul}},\
  }\href
  {http://stacks.iop.org/0953-8984/10/i=44/a=002?key=crossref.5d47e197dfd2b29d12f04d9521e03d5a}
  {\bibfield  {journal} {\bibinfo  {journal} {J. Phys.: Condens. Mat.}\
  }\textbf {\bibinfo {volume} {10}},\ \bibinfo {pages} {L723} (\bibinfo {year}
  {1998})}\BibitemShut {NoStop}%
\bibitem [{\citenamefont {K{\"u}chler}\ \emph {et~al.}(2012)\citenamefont
  {K{\"u}chler}, \citenamefont {Bauer}, \citenamefont {Brando},\ and\
  \citenamefont {Steglich}}]{kuchler_compact_2012}%
  \BibitemOpen
  \bibfield  {author} {\bibinfo {author} {\bibfnamefont {R.}~\bibnamefont
  {K{\"u}chler}}, \bibinfo {author} {\bibfnamefont {T.}~\bibnamefont {Bauer}},
  \bibinfo {author} {\bibfnamefont {M.}~\bibnamefont {Brando}}, \ and\ \bibinfo
  {author} {\bibfnamefont {F.}~\bibnamefont {Steglich}},\ }\href
  {http://aip.scitation.org/doi/10.1063/1.4748864} {\bibfield  {journal}
  {\bibinfo  {journal} {Rev. Sci. Instrum.}\ }\textbf {\bibinfo {volume}
  {83}},\ \bibinfo {pages} {095102} (\bibinfo {year} {2012})}\BibitemShut
  {NoStop}%
\bibitem [{\citenamefont {Rotter}\ \emph {et~al.}(2000)\citenamefont {Rotter},
  \citenamefont {Le}, \citenamefont {Keller}, \citenamefont {Pascut},
  \citenamefont {Hoffmann}, \citenamefont {Doerr}, \citenamefont {Schedler},
  \citenamefont {Fabi}, \citenamefont {Rotter},\ and\ \citenamefont
  {Banks}}]{rotter_mcphase_nodate}%
  \BibitemOpen
  \bibfield  {author} {\bibinfo {author} {\bibfnamefont {M.}~\bibnamefont
  {Rotter}}, \bibinfo {author} {\bibfnamefont {D.~M.}\ \bibnamefont {Le}},
  \bibinfo {author} {\bibfnamefont {J.}~\bibnamefont {Keller}}, \bibinfo
  {author} {\bibfnamefont {L.~G.}\ \bibnamefont {Pascut}}, \bibinfo {author}
  {\bibfnamefont {T.}~\bibnamefont {Hoffmann}}, \bibinfo {author}
  {\bibfnamefont {M.}~\bibnamefont {Doerr}}, \bibinfo {author} {\bibfnamefont
  {R.}~\bibnamefont {Schedler}}, \bibinfo {author} {\bibfnamefont
  {P.}~\bibnamefont {Fabi}}, \bibinfo {author} {\bibfnamefont {S.}~\bibnamefont
  {Rotter}}, \ and\ \bibinfo {author} {\bibfnamefont {M.}~\bibnamefont
  {Banks}},\ }\href {www.mcphase.de} {\enquote {\bibinfo {title} {{McPhase}
  {Software} {Suite}},}\ } (\bibinfo {year} {2000})\BibitemShut {NoStop}%
\bibitem [{\citenamefont {Hiroi}\ \emph {et~al.}(2003)\citenamefont {Hiroi},
  \citenamefont {Matsuhira}, \citenamefont {Takagi}, \citenamefont {Tayama},\
  and\ \citenamefont {Sakakibara}}]{hiroi_specific_2003}%
  \BibitemOpen
  \bibfield  {author} {\bibinfo {author} {\bibfnamefont {Z.}~\bibnamefont
  {Hiroi}}, \bibinfo {author} {\bibfnamefont {K.}~\bibnamefont {Matsuhira}},
  \bibinfo {author} {\bibfnamefont {S.}~\bibnamefont {Takagi}}, \bibinfo
  {author} {\bibfnamefont {T.}~\bibnamefont {Tayama}}, \ and\ \bibinfo {author}
  {\bibfnamefont {T.}~\bibnamefont {Sakakibara}},\ }\href
  {http://journals.jps.jp/doi/abs/10.1143/JPSJ.72.411} {\bibfield  {journal}
  {\bibinfo  {journal} {J. Phys. Soc. Jpn.}\ }\textbf {\bibinfo {volume}
  {72}},\ \bibinfo {pages} {411} (\bibinfo {year} {2003})}\BibitemShut
  {NoStop}%
\bibitem [{\citenamefont {Revell}\ \emph {et~al.}(2013)\citenamefont {Revell},
  \citenamefont {Yaraskavitch}, \citenamefont {Mason}, \citenamefont {Ross},
  \citenamefont {Noad}, \citenamefont {Dabkowska}, \citenamefont {Gaulin},
  \citenamefont {Henelius},\ and\ \citenamefont
  {Kycia}}]{revell_evidence_2013}%
  \BibitemOpen
  \bibfield  {author} {\bibinfo {author} {\bibfnamefont {H.~M.}\ \bibnamefont
  {Revell}}, \bibinfo {author} {\bibfnamefont {L.~R.}\ \bibnamefont
  {Yaraskavitch}}, \bibinfo {author} {\bibfnamefont {J.~D.}\ \bibnamefont
  {Mason}}, \bibinfo {author} {\bibfnamefont {K.~A.}\ \bibnamefont {Ross}},
  \bibinfo {author} {\bibfnamefont {H.~M.~L.}\ \bibnamefont {Noad}}, \bibinfo
  {author} {\bibfnamefont {H.~A.}\ \bibnamefont {Dabkowska}}, \bibinfo {author}
  {\bibfnamefont {B.~D.}\ \bibnamefont {Gaulin}}, \bibinfo {author}
  {\bibfnamefont {P.}~\bibnamefont {Henelius}}, \ and\ \bibinfo {author}
  {\bibfnamefont {J.~B.}\ \bibnamefont {Kycia}},\ }\href
  {http://www.nature.com/articles/nphys2466} {\bibfield  {journal} {\bibinfo
  {journal} {Nat. Phys.}\ }\textbf {\bibinfo {volume} {9}},\ \bibinfo {pages}
  {34} (\bibinfo {year} {2013})}\BibitemShut {NoStop}%
\bibitem [{\citenamefont {Fukazawa}\ \emph {et~al.}(2002)\citenamefont
  {Fukazawa}, \citenamefont {Melko}, \citenamefont {Higashinaka}, \citenamefont
  {Maeno},\ and\ \citenamefont {Gingras}}]{fukazawa_magnetic_2002}%
  \BibitemOpen
  \bibfield  {author} {\bibinfo {author} {\bibfnamefont {H.}~\bibnamefont
  {Fukazawa}}, \bibinfo {author} {\bibfnamefont {R.}~\bibnamefont {Melko}},
  \bibinfo {author} {\bibfnamefont {R.}~\bibnamefont {Higashinaka}}, \bibinfo
  {author} {\bibfnamefont {Y.}~\bibnamefont {Maeno}}, \ and\ \bibinfo {author}
  {\bibfnamefont {M.}~\bibnamefont {Gingras}},\ }\href
  {http://link.aps.org/doi/10.1103/PhysRevB.65.054410} {\bibfield  {journal}
  {\bibinfo  {journal} {Phys. Rev. B}\ }\textbf {\bibinfo {volume} {65}},\
  \bibinfo {pages} {054410} (\bibinfo {year} {2002})}\BibitemShut {NoStop}%
\bibitem [{\citenamefont {Higashinaka}\ \emph {et~al.}(2004)\citenamefont
  {Higashinaka}, \citenamefont {Fukazawa}, \citenamefont {Deguchi},\ and\
  \citenamefont {Maeno}}]{higashinaka_kagome_2004}%
  \BibitemOpen
  \bibfield  {author} {\bibinfo {author} {\bibfnamefont {R.}~\bibnamefont
  {Higashinaka}}, \bibinfo {author} {\bibfnamefont {H.}~\bibnamefont
  {Fukazawa}}, \bibinfo {author} {\bibfnamefont {K.}~\bibnamefont {Deguchi}}, \
  and\ \bibinfo {author} {\bibfnamefont {Y.}~\bibnamefont {Maeno}},\ }\href
  {http://stacks.iop.org/0953-8984/16/i=11/a=015?key=crossref.be294edff9f90badc1934b79f2d29c52}
  {\bibfield  {journal} {\bibinfo  {journal} {J. Phys.: Condens. Mat.}\
  }\textbf {\bibinfo {volume} {16}},\ \bibinfo {pages} {S679} (\bibinfo {year}
  {2004})}\BibitemShut {NoStop}%
\bibitem [{\citenamefont {Sakakibara}\ \emph {et~al.}(2003)\citenamefont
  {Sakakibara}, \citenamefont {Tayama}, \citenamefont {Hiroi}, \citenamefont
  {Matsuhira},\ and\ \citenamefont {Takagi}}]{sakakibara_observation_2003}%
  \BibitemOpen
  \bibfield  {author} {\bibinfo {author} {\bibfnamefont {T.}~\bibnamefont
  {Sakakibara}}, \bibinfo {author} {\bibfnamefont {T.}~\bibnamefont {Tayama}},
  \bibinfo {author} {\bibfnamefont {Z.}~\bibnamefont {Hiroi}}, \bibinfo
  {author} {\bibfnamefont {K.}~\bibnamefont {Matsuhira}}, \ and\ \bibinfo
  {author} {\bibfnamefont {S.}~\bibnamefont {Takagi}},\ }\href
  {http://link.aps.org/doi/10.1103/PhysRevLett.90.207205} {\bibfield  {journal}
  {\bibinfo  {journal} {Phys. Rev. Lett.}\ }\textbf {\bibinfo {volume} {90}},\
  \bibinfo {pages} {207205} (\bibinfo {year} {2003})}\BibitemShut {NoStop}%
\bibitem [{\citenamefont {Bloxsom}(2016)}]{bloxsom_thermal_2016}%
  \BibitemOpen
  \bibfield  {author} {\bibinfo {author} {\bibfnamefont {J.~A.}\ \bibnamefont
  {Bloxsom}},\ }\emph {\bibinfo {title} {Thermal and Magnetic Studies of Spin
  Ice Compounds}},\ \href {http://discovery.ucl.ac.uk/1529383/} {Ph.D.
  thesis},\ \bibinfo  {school} {University College London}, \bibinfo {address}
  {London} (\bibinfo {year} {2016})\BibitemShut {NoStop}%
\bibitem [{\citenamefont {Giblin}\ \emph
  {et~al.}(2018{\natexlab{b}})\citenamefont {Giblin}, \citenamefont
  {Twengstr{\"o}m}, \citenamefont {Bovo}, \citenamefont {Ruminy}, \citenamefont
  {Bartkowiak}, \citenamefont {Manuel}, \citenamefont {Andresen}, \citenamefont
  {Prabhakaran}, \citenamefont {Balakrishnan}, \citenamefont {Pomjakushina},
  \citenamefont {Paulsen}, \citenamefont {Lhotel}, \citenamefont {Keller},
  \citenamefont {Frontzek}, \citenamefont {Capelli}, \citenamefont {Zaharko},
  \citenamefont {McClarty}, \citenamefont {Bramwell}, \citenamefont
  {Henelius},\ and\ \citenamefont {Fennell}}]{giblin_pauling_2018}%
  \BibitemOpen
  \bibfield  {author} {\bibinfo {author} {\bibfnamefont {S.~R.}\ \bibnamefont
  {Giblin}}, \bibinfo {author} {\bibfnamefont {M.}~\bibnamefont
  {Twengstr{\"o}m}}, \bibinfo {author} {\bibfnamefont {L.}~\bibnamefont
  {Bovo}}, \bibinfo {author} {\bibfnamefont {M.}~\bibnamefont {Ruminy}},
  \bibinfo {author} {\bibfnamefont {M.}~\bibnamefont {Bartkowiak}}, \bibinfo
  {author} {\bibfnamefont {P.}~\bibnamefont {Manuel}}, \bibinfo {author}
  {\bibfnamefont {J.~C.}\ \bibnamefont {Andresen}}, \bibinfo {author}
  {\bibfnamefont {D.}~\bibnamefont {Prabhakaran}}, \bibinfo {author}
  {\bibfnamefont {G.}~\bibnamefont {Balakrishnan}}, \bibinfo {author}
  {\bibfnamefont {E.}~\bibnamefont {Pomjakushina}}, \bibinfo {author}
  {\bibfnamefont {C.}~\bibnamefont {Paulsen}}, \bibinfo {author} {\bibfnamefont
  {E.}~\bibnamefont {Lhotel}}, \bibinfo {author} {\bibfnamefont
  {L.}~\bibnamefont {Keller}}, \bibinfo {author} {\bibfnamefont
  {M.}~\bibnamefont {Frontzek}}, \bibinfo {author} {\bibfnamefont {S.~C.}\
  \bibnamefont {Capelli}}, \bibinfo {author} {\bibfnamefont {O.}~\bibnamefont
  {Zaharko}}, \bibinfo {author} {\bibfnamefont {P.~A.}\ \bibnamefont
  {McClarty}}, \bibinfo {author} {\bibfnamefont {S.~T.}\ \bibnamefont
  {Bramwell}}, \bibinfo {author} {\bibfnamefont {P.}~\bibnamefont {Henelius}},
  \ and\ \bibinfo {author} {\bibfnamefont {T.}~\bibnamefont {Fennell}},\ }\href
  {https://link.aps.org/doi/10.1103/PhysRevLett.121.067202} {\bibfield
  {journal} {\bibinfo  {journal} {Phys. Rev. Lett.}\ }\textbf {\bibinfo
  {volume} {121}},\ \bibinfo {pages} {067202} (\bibinfo {year}
  {2018}{\natexlab{b}})}\BibitemShut {NoStop}%
\bibitem [{\citenamefont {Henelius}\ \emph {et~al.}(2016)\citenamefont
  {Henelius}, \citenamefont {Lin}, \citenamefont {Enjalran}, \citenamefont
  {Hao}, \citenamefont {Rau}, \citenamefont {Altosaar}, \citenamefont
  {Flicker}, \citenamefont {Yavors'kii},\ and\ \citenamefont
  {Gingras}}]{henelius_refrustration_2016}%
  \BibitemOpen
  \bibfield  {author} {\bibinfo {author} {\bibfnamefont {P.}~\bibnamefont
  {Henelius}}, \bibinfo {author} {\bibfnamefont {T.}~\bibnamefont {Lin}},
  \bibinfo {author} {\bibfnamefont {M.}~\bibnamefont {Enjalran}}, \bibinfo
  {author} {\bibfnamefont {Z.}~\bibnamefont {Hao}}, \bibinfo {author}
  {\bibfnamefont {J.~G.}\ \bibnamefont {Rau}}, \bibinfo {author} {\bibfnamefont
  {J.}~\bibnamefont {Altosaar}}, \bibinfo {author} {\bibfnamefont
  {F.}~\bibnamefont {Flicker}}, \bibinfo {author} {\bibfnamefont
  {T.}~\bibnamefont {Yavors'kii}}, \ and\ \bibinfo {author} {\bibfnamefont
  {M.~J.~P.}\ \bibnamefont {Gingras}},\ }\href
  {http://link.aps.org/doi/10.1103/PhysRevB.93.024402} {\bibfield  {journal}
  {\bibinfo  {journal} {Phys. Rev. B}\ }\textbf {\bibinfo {volume} {93}},\
  \bibinfo {pages} {024402} (\bibinfo {year} {2016})}\BibitemShut {NoStop}%
\end{thebibliography}%

\end{document}